\documentclass[reprint,twocolumn,pre,showpacs,amsmath,amssymb,aps]{revtex4-1}

\usepackage{amsmath}
\usepackage{natbib}	
\usepackage{graphicx,color,rotating}
\usepackage[latin1]{inputenc}
\usepackage{textcomp}
\usepackage{dcolumn}
\usepackage{bm}     
\usepackage{upgreek}
\usepackage{subdepth}  			
\usepackage[latin1]{inputenc}

\usepackage{array}
\newcolumntype{L}[1]{>{\raggedright\let\newline\\\arraybackslash\hspace{0pt}}m{#1}}
\newcolumntype{C}[1]{>{\centering\let\newline\\\arraybackslash\hspace{0pt}}m{#1}}
\newcolumntype{R}[1]{>{\raggedleft\let\newline\\\arraybackslash\hspace{0pt}}m{#1}}

\usepackage{hyperref}						
\usepackage{xcolor}						
\hypersetup{							
    colorlinks,
    linkcolor={red!50!black},
    citecolor={green!30!black},
    urlcolor={blue!80!black}
}
\newcommand{\mr}[1]{\ensuremath{\mathrm{#1}}}
\renewcommand{\vec}[1]{\bm{#1}}
\newcommand{\ee}{\mathrm{e}}
\newcommand{\ii}{\mathrm{i}}
\newcommand{\dm}{\mathrm{d}}
\newcommand{\avr}[1]{\big\langle #1 \big\rangle}

\newcommand{\taut}{\boldsymbol{\tau}}

\newcommand{\pp}{\partial}
\newcommand{\Pdiss}{P_\mr{visc}^\mr{diss}}
\newcommand{\Ploss}{P_\mr{loss}}
\newcommand{\Pvisc}{P_\mr{visc}}
\newcommand{\Psurr}{P_\mr{visc}^\mr{wall}}

\newcommand{\nablabf}{\boldsymbol{\nabla}}

\newcommand{\Lapl}{\nabla^2}

\newcommand{\rot}{\nablabf\times}

\newcommand{\ppt}{\partial_t}

\newcommand{\etal}{\textit{et~al.}}

\newcommand{\scap}{\!\cdot\!}
\newcommand{\scapmat}{\!:\!}

\newcommand{\AAA}{\vec{A}}

\newcommand{\BBB}{\vec{B}}

\newcommand{\eee}{\vec{e}}
\newcommand{\een}{\vec{e}}

\newcommand{\Imat}{\textbf{\textsf{I}}}

\newcommand{\kc}{k_\mathrm{c}}

\newcommand{\ks}{k_\mathrm{s}}

\newcommand{\nnn}{\vec{n}}

\newcommand{\rrr}{\vec{r}}

\newcommand{\SSS}{\vec{S}}
\newcommand{\SSSac}{\vec{S}_\mathrm{ac}^d}
\newcommand{\sss}{\vec{s}}

\newcommand{\uuu}{\vec{u}}

\newcommand{\VVV}{\vec{V}}

\newcommand{\vvv}{\vec{v}}

\newcommand{\zerovec}{\boldsymbol{0}}

\newcommand{\cL}{c_\mathrm{lo}}

\newcommand{\cO}{c_0}

\newcommand{\cT}{c_\mathrm{tr}}

\newcommand{\Eac}{E_\mathrm{ac}}
\newcommand{\Eackin}{E_\mr{ac}^\mr{kin}}
\newcommand{\Eacpot}{E_\mr{ac}^\mr{pot}}
\newcommand{\Eacbar}{\bar{E}^d_\mathrm{ac}}

\newcommand{\Eacbarres}{\bar{E}^{d,\mr{res}}_\mathrm{ac}}

\newcommand{\Lac}{\mathcal{L}_\mathrm{ac}}

\renewcommand{\perp}{\zeta}

\newcommand{\eps}{\epsilon}

\newcommand{\etaBO}{\eta^\mathrm{b}_0}

\newcommand{\etaO}{\eta_0}

\newcommand{\Gambl}{\Gamma_\mathrm{bl}}

\newcommand{\nuO}{\nu_0}

\newcommand{\Omgf}{{\Omega_\mr{fl}}}
\newcommand{\Omgs}{{\Omega_\mr{sl}}}

\newcommand{\ppti}{\tilde{\partial}}

\newcommand{\yti}{{\tilde{y}{}}}
\newcommand{\zti}{{\tilde{z}{}}}

\newcommand{\cOsqr}{c^{\,2_{}}_0}

\newcommand{\fres}{f_\mathrm{res}}
\newcommand{\omegares}{\omega_\mathrm{res}}

\newcommand{\kapO}{\kappa_0}

\newcommand{\pI}{p_1}

\newcommand{\vvvI}{\vvv_1}

\newcommand{\rhoO}{\rho_0}

\newcommand{\rhoI}{\rho_1}

\newcommand{\SImum}{\textrm{\textmu{}m}}

\newcommand{\SInm}{\textrm{nm}}

\newcommand{\nn}{\nonumber}
\newcommand{\beq}[1]{\begin{equation} \eqlab{#1}}
\newcommand{\eeq}{\end{equation}}
\newcommand{\bsub}{\begin{subequations}}
\newcommand{\esub}{\end{subequations}}
\def\bal#1\eal{\begin{align}#1\end{align}}
\def\balat#1#2\ealat{\begin{alignat}{#1} #2 \end{alignat}}
\def\bsubal#1 #2\esubal{\bsuba{#1}\begin{align}#2\end{align} \esuba}     
\def\bsubalat#1#2\esubalat{\bsub \begin{alignat}{#1} #2 \end{alignat} \esub}
\newcommand{\bsuba}[1]{\bsub \eqlab{#1}}
\newcommand{\esuba}{\esub}

\newcommand{\eqlab}[1]{\label{eq:#1}}
\renewcommand{\eqref}[1]{Eq.~(\ref{eq:#1})}
\newcommand{\eqnoref}[1]{(\ref{eq:#1})}

\newcommand{\eqsref}[2]{Eqs.~(\ref{eq:#1}) and~(\ref{eq:#2})}
\newcommand{\eqsnoref}[2]{(\ref{eq:#1}) and~(\ref{eq:#2})}

\newcommand{\figref}[1]{Fig.~\ref{fig:#1}}

\newcommand{\figlab}[1]{\label{fig:#1}}
\newcommand{\appref}[1]{Appendix~\ref{sec:#1}}

\newcommand{\secref}[1]{Section~\ref{sec:#1}}
\newcommand{\secsref}[2]{Sections~\ref{sec:#1} and~\ref{sec:#2}}
\newcommand{\secsrangeref}[2]{Sections~\ref{sec:#1}--\ref{sec:#2}}
\newcommand{\seclab}[1]{\label{sec:#1}}
\newcommand{\tabref}[1]{Table~\ref{tab:#1}}

\newcommand{\tablab}[1]{\label{tab:#1}}

\newcommand{\abs}[1]{\left|{#1}\right|}
\newcommand{\sigmat}{\boldsymbol{\sigma}}

\newcommand{\ord}[1]{\mathcal{O}({#1})}
\newcommand{\eikOx}{\ee^{\ii k_0 x}}

\newcommand{\eiot}{\ee^{-\ii\omega t}}

\newcommand{\rhoS}{\rho_\mathrm{s}}
\newcommand{\sigmatS}{\sigmat_\mathrm{s}}
\newcommand{\GammaS}{\Gamma_\mathrm{s}}

\newcommand{\xs}{\xi}
\newcommand{\ys}{\eta}
\newcommand{\zs}{\zeta}
\renewcommand{\H}{\mathcal{H}}

\newcommand{\ex}{\eee_{x}}

\newcommand{\exs}{\eee_{\xs}}
\newcommand{\eys}{\eee_{\ys}}
\newcommand{\ezs}{\eee_{\zs}}

\newcommand{\grad}{\boldsymbol{\nabla}}
\newcommand{\pargrad}{\boldsymbol{\nabla}_\parallel}
\newcommand{\perpgrad}{\boldsymbol{\nabla}_\perp}
\renewcommand{\div}{\nablabf\!\cdot}
\newcommand{\pardiv}{\nablabf_{\parallel}^{}\scap}
\newcommand{\perpdiv}{\nablabf_{\perp}^{}\scap}
\newcommand{\lap}{\nabla^2}

\newcommand{\pdiff}[2]{\dfrac{\partial{#1}}{\partial{#2}}}

\newcommand{\intI}[2]{I_{{#1}{#2}}^{(1)}}
\newcommand{\intII}[2]{I_{{#1}{#2}}^{(2)}}
\newcommand{\intIII}[2]{I_{{#1}{#2}}^{(3)}}
\newcommand{\intn}[2]{I_{{#1}{#2}}^{(n)}}
\renewcommand{\Re}{\mathrm{Re}}

\newcommand{\shortrange}{\delta} 			
\newcommand{\longrange}{d}				
\newcommand{\atsurface}{0}				

\newcommand{\vshort}{v}				
\newcommand{\vvvshort}{\vvv}
\newcommand{\vlong}{v}					
\newcommand{\vvvlong}{\vvv}				
\newcommand{\pshort}{p}				
\newcommand{\plong}{p}					
\newcommand{\vvvwall}{\VVV^\atsurface}				
\newcommand{\vvvwallC}{\VVV^{\atsurface *}}		    
\newcommand{\vvvwallpar}{\vvvwall_{1\parallel}}

\newcommand{\vwall}{V^\atsurface}					
\newcommand{\vwallC}{V^{\atsurface *}}				
\newcommand{\vwallperp}{\vwall_{1\perp}}
\newcommand{\vwallperpC}{\vwallC_{1\perp}}
\newcommand{\g}{g}						
\newcommand{\gshort}{g}				

\newcommand{\gd}{\gshort^{\shortrange}}

\newcommand{\AAAs}{\AAA^\atsurface}
\newcommand{\BBBs}{\BBB^\atsurface}

\newcommand{\AAAspar}{\AAA_\parallel^\atsurface}

\newcommand{\Asperp}{A_\perp^\atsurface}
\newcommand{\As}{A^\atsurface}
\newcommand{\Bs}{B^\atsurface}

\newcommand{\vd}{\vshort^\shortrange}
\newcommand{\vds}{\vshort^{\shortrange\atsurface}}
\newcommand{\vdsC}{\vshort^{\shortrange\atsurface *}}

\newcommand{\vdsperp}{\vshort^{\shortrange \atsurface}_{1\perp}}
\newcommand{\vdsCperp}{\vshort^{\shortrange \atsurface*}_{1\perp}}
\newcommand{\vvvd}{\vvvshort^\shortrange}
\newcommand{\vvvds}{\vvvshort^{\shortrange\atsurface}}
\newcommand{\vvvdsC}{\vvvshort^{\shortrange\atsurface*}}
\newcommand{\vvvdspar}{\vvvshort^{\shortrange\atsurface}_{1\parallel}}

\newcommand{\sigmatd}{\sigmat^\shortrange}
\newcommand{\sigmatl}{\sigmat^\longrange}

\newcommand{\pd}{\pshort^{\shortrange}}

\newcommand{\vl}{\vlong^\longrange}
\newcommand{\vlC}{\vlong^{\longrange *}}
\newcommand{\vls}{\vlong^{\longrange\atsurface}}
\newcommand{\vlsC}{\vlong^{\longrange\atsurface *}}

\newcommand{\vlperp}{\vlong^\longrange_{1\perp}}
\newcommand{\vlCperp}{\vlong^{\longrange*}_{1\perp}}
\newcommand{\vlsperp}{\vlong^{\longrange\atsurface}_{1\perp}}
\newcommand{\vlsCperp}{\vlong^{\longrange\atsurface*}_{1\perp}}

\newcommand{\vvvl}{\vvvlong^\longrange}
\newcommand{\vvvlC}{\vvvlong^{\longrange *}}
\newcommand{\vvvls}{\vvvlong^{\longrange\atsurface}}
\newcommand{\vvvlsC}{\vvvlong^{\longrange\atsurface*}}

\newcommand{\vvvlspar}{\vvvlong^{\longrange\atsurface}_{1\parallel}}

\renewcommand{\pl}{\plong^{\longrange}}

\begin{document}

\title{Theory of pressure acoustics with boundary layers\\ and streaming in curved elastic cavities}

\author{Jacob S. Bach}
\email{jasoba@fysik.dtu.dk}
\affiliation{Department of Physics, Technical University of Denmark,\\ DTU Physics Building 309, DK-2800 Kongens Lyngby, Denmark}

\author{Henrik Bruus}
\email{bruus@fysik.dtu.dk}
\affiliation{Department of Physics, Technical University of Denmark,\\
DTU Physics Building 309, DK-2800 Kongens Lyngby, Denmark}

\date{16 April 2018}

\begin{abstract}
The acoustic fields and streaming in a confined fluid depend strongly on the acoustic boundary layer forming near the wall. The width of this layer is typically much smaller than the bulk length scale set by the geometry or the acoustic wavelength, which makes direct numerical simulations challenging. Based on this separation in length scales, we extend the classical theory of pressure acoustics by deriving a boundary condition for the acoustic pressure that takes boundary-layer effects fully into account. Using the same length-scale separation for the steady second-order streaming, and combining it with time-averaged short-range products of first-order fields, we replace the usual limiting-velocity theory with an analytical slip-velocity condition on the long-range streaming field at the wall. The derived boundary conditions are valid for oscillating cavities of arbitrary shape and wall motion as long as the wall curvature and displacement amplitude are both sufficiently small. Finally, we validate our theory by comparison with direct numerical simulation in two examples of two-dimensional water-filled cavities: The well-studied rectangular cavity with prescribed wall actuation, and the more generic elliptical cavity embedded in an externally actuated rectangular elastic glass block.
\end{abstract}




\maketitle


\section{Introduction}
The study of ultrasound effects in fluids in sub-millimeter cavities and channels has intensified the past decade, as microscale acoustofluidic devices are used increasingly in biology, environmental and forensic sciences, and clinical diagnostics \citep{Bruus2011c, Laurell2014}. Examples include cell synchronization \citep{Thevoz2010}, enrichment of prostate cancer cells in blood \citep{Augustsson2012}, size-independent sorting of cells \citep{Augustsson2016}, manipulation of \textit{C. elegans} \citep{Ding2012}, and single-cell patterning~\citep{Collins2015}. Acoustics can also be used for non-contact microfluidic trapping and particle enrichment \citep{Hammarstrom2010, Hammarstrom2012, Hammarstrom2014} as well as acoustic tweezing  \citep{Drinkwater2016, Collins2016, Lim2016, Baresch2016}.

The two fundamental physical phenomena that enable these microscale acoustofluidic applications are rooted in nonlinear acoustics. One fundamental phenomenon is the acoustic radiation force, which tends to focus suspended particles in pressure nodes based on their acoustic contrast to the surrounding fluid \citep{King1934, Yosioka1955, Doinikov1997, Doinikov1997a, Doinikov1997b, Settnes2012, Karlsen2015}. The second fundamental phenomenon is the acoustic streaming appearing as steady flow rolls which tend to defocus suspended particles due to Stokes drag \citep{LordRayleigh1884, Schlichting1932, Wiklund2012, Muller2013, Lei2016, Riaud2017}. Because the acoustic radiation force scales with the volume of the suspended particle, and the Stokes drag with its radius, the former dominates for large particles and the latter for small. For water at room temperature and 1~MHz ultrasound, the critical particle radius for the crossover between these two regimes has been determined to be around $2~\SImum$ \citep{Muller2012, Barnkob2012a}.

So far, the vast majority of successful microscale acoustofluidics applications has been for large (above $2~\SImum$) particles, such as cells, whose dynamics is dominated by the well-characterized, robust acoustic radiation force, which depends on the bulk properties of the acoustic field and material parameters of the particles and the surrounding fluid. However, there is a strong motivation to handle also sub-micrometer particles such as bacteria, exosomes, and viruses, for use in contemporary lab-on-a-chip-based diagnostics and biomedical research \citep{Hammarstrom2012, Antfolk2014, Mao2017, Wu2017}. In contrast to large particles, the dynamics of small (sub-micrometer) particles is dominated by the drag force from the ill-characterized acoustic streaming, and because this streaming is partly driven by the Reynolds stress in the sub-micrometer-thin acoustic boundary layers, it becomes highly sensitive to details of the geometry, motion, and temperature of the confining oscillating walls. To control the handling of such nanoparticle suspensions, a deeper understanding of the often complicated acoustic streaming is called for.

One important aspect of ultrasound acoustics is the large velocity gradients in the acoustic boundary layer near rigid boundaries \citep{LordRayleigh1884}. The Reynolds stress building up in this region is responsible for both the viscous damping of the harmonic acoustic fields and for the generation of time-averaged momentum flux giving rise to acoustic streaming. In water with kinematic viscosity $\nu_0\approx  10^{-6} \ \mathrm{m}^2/\mathrm{s}$ at the frequency $f=\frac{1}{2\pi}\omega\approx 1$~MHz, the thickness $\delta = \sqrt{2\nu_0/\omega}$ of this boundary layer is of the order of 500~nm, while the acoustic wavelength is around 1.5~mm. This three-orders-of-magnitude separation of physically relevant length scales poses a severe challenge for numerical simulations. To circumvent the problem of resolving the thin boundary layer, we develop a theory for pressure acoustics with boundary-layers and streaming that allows calculations of the pressure field and bulk streaming field which both varies on the much longer length scale $d\gg \delta$.

First, we extend the classical pressure acoustics theory by formulating a boundary condition for the acoustic pressure that includes the presence of the boundary layer, which is otherwise neglected. Thus, our extended boundary condition takes into account important effect of the boundary layer, such as increased viscous damping, shifts in resonance frequencies, and shear stresses on the surrounding walls.

Second, we formulate a generalized slip boundary condition for bulk acoustic streaming over curved oscillating surfaces. An important step in this direction was the development of the limiting-velocity theory by Nyborg in 1958~\citep{Nyborg1958} for perpendicularly oscillating curved walls. Later modifications of this theory comprise modifications to the analysis in curvilinear coordinates by Lee and Wang in 1989~\citep{Lee1989}, and the treatment of oscillations in any direction for flat walls by Vanneste and B\"{u}hler in 2011~\citep{Vanneste2011}. Here, we extend these theories to harmonic oscillations in any direction of an arbitrarily shaped, elastic wall provided that both the radius of curvature and the acoustic wavelength are much larger than the boundary layer length-scale $\delta$, and that also the amplitude of the perpendicular surface vibration is much smaller than $\delta$.

Notably, the theoretical description developed here allows us to perform numerical simulations of the linear and nonlinear acoustics in arbitrarily shaped liquid-filled cavities embedded in oscillating elastic solids. Examples and validation of such simulations for two-dimensional (2D) systems are presented in the final sections of this paper, while a study of three-dimensional (3D) systems is work in progress to be presented later.

\section{Wall motion and perturbation theory}

We consider a fluid domain $\Omega$ bounded by an elastic, oscillating solid, see \figref{surface_sketch}. All acoustic effects in the fluid are generated by the fluid-solid interface that oscillates harmonically around its equilibrium position, denoted $\sss_0$ or $\pp \Omega$, with an angular frequency $\omega$. The instantaneous position $\sss(\sss_0,t)$ at time $t$ of this interface (the wall), is described by the small complex displacement $\sss_1(\sss_0)\eiot$,
 \beq{sss}
 \sss(\sss_0,t)=\sss_0+\sss_1(\sss_0)\:\eiot.
 \eeq
In contrast to Muller and Bruus~\citep{Muller2015}, we do not study the transient phase leading to this steady oscillatory motion.

 \begin{figure}[!t]
 \centering
 \includegraphics[width=\columnwidth]{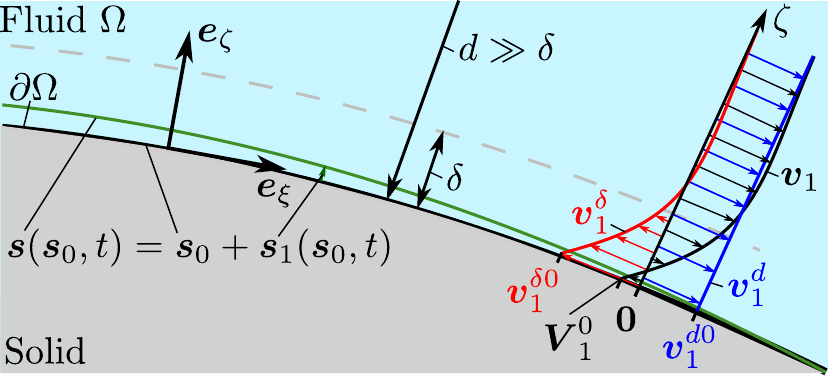}
 \caption[]{\figlab{surface_sketch}
Sketch of the interface between a fluid (light blue, $\Omega$) and a curved, oscillating solid (gray) with instantaneous position  $\sss$ (green line) and equilibrium position $\sss_0$ (black line, $\pp\Omega$). The local curvilinear coordinate system on the interface is given by the tangent vectors $\exs$ and $\eys$ and the normal vector $\ezs$. By a Helmholtz decomposition, the first-order acoustic fluid velocity $\vvv_1 = \vvvl_1+\vvvd_1$ is written as the sum of a long-range compressible part $\vvvl_1$ (blue) extending into the bulk and a short-range incompressible part $\vvvd_1$ (red) with a decay length equal to the boundary-layer width $\delta$. $\vvvwall_1 = \vvvls_1+\vvvds_1$ is the Lagrangian velocity of the interface (the wall).}
 \end{figure}

\subsection{Fundamental conservation laws in acoustofluidics}
The theory of acoustofluidics in $\Omega$ is derived from the conservation of the fluid mass and momentum density,
 \bsub
 \eqlab{cont_navier}
 \bal
 \eqlab{cont}
 \ppt\rho	&=-\div(\rho\vvv),
 \\
 \eqlab{navier}
 \ppt(\rho\vvv)&= -\div[(\rho \vvv)\vvv] + \div\sigmat,
 \eal
where $\rho$ is the mass density, $\vvv$ is the Eulerian fluid velocity, and $\sigmat$ is the viscous stress tensor,  given by
 \bal
 \eqlab{sigma}
 \sigmat &= -p\, \Imat + \taut,
 \\ \eqlab{tau}
 \taut &=\etaBO(\div\vvv)\Imat
 + \etaO \Big[\grad \vvv + (\grad \vvv)^\mathrm{T}-\frac{2}{3}(\div\vvv) \Imat\Big].
 \eal
 \esub
Here, $p$ is the pressure, and $\taut$ is the viscous part of the stress tensor given in terms of the bulk viscosity $\etaBO$, the dynamic viscosity $\eta_0$, the identity matrix $\Imat$, and the superscript "T" denoting matrix transpose. We introduce the isentropic compressibility $\kapO$ and speed of sound $c_0$,
 \beq{kap0}
 \kapO  = \dfrac{1}{\rho_0} \left(\pdiff{\rho}{p}\right)_S = \frac{1}{\rhoO\cOsqr},
 \eeq
as well as the dimensionless damping coefficient $\Gamma$ in terms of the viscosity ratio $\beta$,
 \beq{Gamma_beta}
 \Gamma = (\beta+1)\etaO\omega\kapO,
 \qquad
 \beta  = \frac{\etaBO}{\etaO} + \frac13.
 \eeq

\subsection{Perturbation expansion}
The linear acoustic response of the system is proportional to the displacement stimulus $\sss_1(\sss_0)\eiot$, and the resulting complex-valued quantities $Q_1(\rrr)\:\eiot$ are called first-order fields with subscript "1". The physical time-dependent quantity $Q^\mr{phys}_{1}(\rrr,t)$ corresponding to $Q_1$ is given by the real part $Q^\mr{phys}_{1}(\rrr,t) = \Re\big[Q_1(\rrr)\:\eiot\big]$.

As the governing equations are nonlinear, we also encounter higher-order terms. In the present work, we only include terms to second order in the stimulus. Moreover, since we are only interested in the steady part of these second-order fields, we let in the following the subscript "2" denote a time-averaged quantity, written as $Q_2(\rrr) = \avr{Q_2(\rrr,t)} = \frac{\omega}{2\pi}\int_0^{2\pi/\omega} Q_2(\rrr,t)\ \dm t$. Time-averages of products of time-harmonic complex-valued first-order fields $A_1$ and $B_1$ are also of second order, and for those we have $\avr{A_1B_1}=\frac12\mathrm{Re}\big[A_1(\rrr) B_1^*(\rrr)\big]$, where the asterisk denote complex conjugation.

Using this notation for the fluid, we expand the mass density $\rho$, the pressure $p$, and the velocity $\vvv$ in perturbation series of the form,
 \bsub
 \eqlab{def_pert_series}
 \balat{4}{\eqlab{def_rho}
 \rho  &= \rho_0	&&\,+\, \rho_1(\rrr)\eiot	
 &&\,+\,\rho_2(\rrr),
 \\
 \eqlab{def_p}
 p &= p_0 		&&\,+\, p_1(\rrr)\eiot 		
 &&\,+\, p_2(\rrr),
 \\
 \eqlab{def_v}
 \vvv &= \zerovec 	&&\,+\, \vvv_1(\rrr)\eiot 		
 &&\,+\, \vvv_2(\rrr),
 }\ealat
 \esub
where $\rhoI \ll \rhoO$, $\pI = \cOsqr\rhoI \ll  \cOsqr\rhoO$, and $\big|\vvvI\big| \ll \cO$. The subscripts 1 and 2 denote the order in the small acoustic Mach number $\mathrm{Ma} = \abs{\vvv_1}/{\cO}$, which itself is proportional to $\sss_1$.

\subsection{No-slip boundary condition at the wall}
To characterize the wall motion, we compute the time derivative of $\sss(\sss_0,t)$ in \eqref{sss},
 \beq{dtsss}
 \pp_t \sss(\sss_0,t)=-\ii\omega \sss_1(\sss_0)\: \eiot = \vvvwall_1(\sss_0)\:\eiot,
 \eeq
where $\vvvwall_1(\sss_0) = -\ii\omega\sss_1(\sss_0)$ is the Lagrangian velocity of the wall surface element with equilibrium position $\sss_0$ and instantaneous position $\sss$. The no-slip boundary condition on the Eulerian fluid velocity $\vvv(\rrr,t)$ is imposed at the instantaneous surface position $\sss(t)$,\citep{Vanneste2011, Bradley1996}
 \bal\eqlab{no-slip}
 \vvv(\sss_0+\sss_1\eiot,t)=\vvvwall_1(\sss_0)\:\eiot,\: \text{no-slip condition. }
 \eal
Combining \eqsref{no-slip}{def_v} with the Taylor expansion
$\vvv_1(\sss_0+\sss_1,t) \approx \vvv_1(\sss_0)\:\eiot + \avr{(\sss_1\scap\grad)\vvv_1(\sss_0)}$, and collecting the terms order by order, gives
 \bsuba{bc}
 \balat{2}{
 \eqlab{bc_first}
 \vvv_1(\sss_0) &= \vvvwall_1(\sss_0),
 && \text{ 1st-order condition},
 \\
 \eqlab{bc_second}
 \vvv_2(\sss_0) &= -\avr{(\sss_1\scap\grad)\vvv_1}\big|^{{}}_{\sss_0},
 && \text{ 2nd-order condition}.
 }\ealat
 \esuba
Note that the expansion, or Stokes drift, in \eqref{bc_second} is valid if the length scale over which $\vvv_1$ varies is much larger than $\abs{\sss_1}$. So we require $|\sss_{1\parallel}| \ll d$ and $| s_{1\perp}| \ll \delta$.

\subsection{Local boundary-layer coordinates} \seclab{Geometry}
In the boundary layer, we introduce the local coordinates $\xs$, $\ys$, and $\zs$. The latter measures distance away from the surface equilibrium position along the surface unit normal vector $\eee_\zs$, while the tangential coordinates $\xs$ and $\ys$ increase in the respective directions of the unit tangent vectors $\eee_\xs$ and $\eee_\ys$, but not necessarily measuring arc length. We define differential-geometric symbols,
 \bsubalat{2}{
 \eqlab{hiHiDef}
 h_i &= \abs{\pp_i\rrr},
 & \quad
 T_{kji} &= \big(\ppti_k \eee_j\big)\scap\eee_i,
 \text{ for } i,j,k = \xi,\eta,\zeta,
 \\
 \eqlab{pptiTkjiDef}
 \ppti_i &= \dfrac{1}{h_i}\pp_i,
 & \H_k &= T_{iki} = \ppti_k \Big[\sum_{i\neq k} \log h_i\Big],
 }\esubalat
and use them to write the following derivatives involving a scalar field $g$ and two vector fields $\AAA$ and $\BBB$ in the local right-handed, orthogonal, curvilinear coordinate system,
 \bsubal{curvi_derivatives}
 \eqlab{curvi_grad}
 \grad  &= \eee_i \ppti_i,\\
 \eqlab{curvi_lap}
 \lap \g &= (\ppti_i\ppti_i + \H_i\ppti_i)\g,\\
 \eqlab{curvi_div}
 \div \AAA &= (\ppti_i+\H_i) A_i,\\
 \eqlab{curvi_condif}
 (\AAA\cdot\grad)\BBB&=A_k\big(\ppti_k B_i + T_{kji}B_j\big)\eee_i,
\esubal
where summation over repeated indices is implied. Note that since $\zs$ measures arc length, we have $h_\zs=1$ and consequently $\ppti_\zs= \pp_\zs$. It is useful to introduce parallel and perpendicular differential operators $\pargrad$ and $\perpgrad$,
 \bsubal{ParPerpDeriv}
 \eqlab{PardivPerpdivDef}
 \pargrad &= \eee_\xs\ppti_\xs + \eee_\ys\ppti_\ys, \qquad
 \perpgrad = \eee_\zs\ppti_\zs,\\
 \eqlab{ParDiv}
 \pardiv \AAA &= (\ppti_\alpha+\H_\alpha) A_\alpha,
\hspace{4mm} \text{ sum over $\alpha = \xs,\ys$},\\
 \eqlab{PerpDiv}
 \perpdiv \AAA &= (\ppti_\zs+\H_\zs) A_\zs,\\
 \eqlab{ParAdv}
 (\AAA\scap\pargrad)\BBB&=A_\alpha\big(\ppti_\alpha B_i
 +  T_{\alpha ji}B_j \big)\eee_i.
\esubal
where repeated Greek index $\alpha$ only sums over $\xs$ and  $\ys$.

\subsection{Surface fields, boundary-layer fields, and bulk fields}

For fluid fields, we distinguish between boundary-layer fields and bulk fields with superscripts $"\delta"$ and $"d"$, respectively, denoting the length scale of the variations in the perpendicular direction $\ezs$ as shown in \figref{surface_sketch}. Here,
 \beq{deltaDef}
 \delta = \sqrt{\frac{2\nuO}{\omega}} = \sqrt{\frac{2\etaO}{\rhoO\omega}} \approx 500~\SInm \text{ (water at 1~MHz)},
 \eeq
is the short, shear length scale of the acoustic boundary layer, while $d$ is the long compressional length scale being the minimum of the local surface curvature length scale $R$ and the inverse wave number $k_0^{-1}=c_0/\omega$ for sound speed $c_0$. We introduce the ratio $\eps$ of these length scales,
 \beq{epsilon}
 \eps = \dfrac{\delta}{d}\ll 1,
\eeq
where the inequality holds if both $k_0\delta \ll 1$ and $\delta/R\ll1$, a condition usually satisfied in microfluidic devices.

The central point in our theory is that we analyze the weakly curved, thin boundary-layer limit $\eps \ll 1$, where derivatives of boundary-layer fields are included only to lowest order in $\eps$. In this limit, several simplifications can be made, which ultimately allows for analytical results. It is useful to decompose a vector $\AAA$ into parallel and perpendicular components $\AAA_\parallel$ and $\AAA_\perp$, respectively,
 \beq{AvecParallelPerp}
 \AAA = \AAA_\parallel + \AAA_\perp,\text{ with }
 \AAA_\perp = (\AAA\cdot\ezs)\:\ezs = A_\perp\:\een_\perp.
 \eeq
The Laplacian of a boundary-layer scalar $\gd$, \eqref{curvi_lap}, and the divergence of a boundary-layer vector $\AAA^\shortrange$, \eqref{curvi_div}, reduce to
 \bsubal{curvi_derivatives_delta}
 \eqlab{curvi_lap_delta}
 \lap \gd &\approx \pp_\perp^2 \gd,
 \\
 \eqlab{curvi_div_delta}
 \div\AAA^\shortrange &\approx \pardiv \AAA_\parallel^\shortrange
 +\pp_\perp A_\perp^\shortrange.
 \esubal
Further reductions are obtained by separating in the perpendicular coordinate $\zeta$,
 \bsub
 \beq{A0a}
 \AAA(\xs,\ys,\zs)=\AAAs(\xs,\ys) a(\zs), \quad \zs \ll d,
 \eeq
for any field $\AAA$ in the fluid boundary layer. Here, superscript $"\atsurface"$ defines a surface field $\AAAs(\xs,\ys) = \AAA(\xs,\ys,0)$, such as the wall velocity $\vvvwall_1$ and the fluid velocity $\vvv^0$ at the wall. Note that a surface field does not have a perpendicular derivative, although it does have a perpendicular component. For surface fields \eqsref{curvi_div}{curvi_condif} become,
 \eqlab{curvi_derivatives_zero}
 \bal
 \eqlab{curvi_derivatives_zero_a}
 \div \AAAs &= \pardiv\AAAspar+ \H_\perp \Asperp ,
 \\
 \eqlab{curvi_derivatives_zero_b}
 (\AAAs\scap\grad)\BBBs&=\big[(\AAAspar\cdot\pargrad)  \Bs_i\big]\eee_i+\As_k\Bs_jT_{kji}\eee_i.
 \eal
 \esub
With this, we have established the necessary notation. In summary, the length-scale conditions for the following boundary-layer theory to be valid are,
 \balat{2} \eqlab{lengthscale_assumptions}
 &\text{Compressional length scale $d=\min\big\{R,k_0^{-1}\big\}$},
 & \delta &\ll d,
 \nn \\
 &\text{Parallel wall displacement $| \sss_{1\parallel}|$ },
 \;\;& | \sss_{1\parallel}| &\ll d,
 \nn \\
 &\text{Perpendicular wall displacement  $|s_{1\perp} |$},
 & |s_{1\perp} | & \ll \delta.
 \ealat

\section{First-order time-harmonic fields }\seclab{first}
\seclab{first_theory}
To first order in $\mathrm{Ma}=\frac{1}{\cO}\abs{\vvv_1}$, Eqs.~\eqnoref{cont_navier} and \eqnoref{def_pert_series} give,
 \bsubal{cont_navier_first}
 \eqlab{p1_rho1_first}
 \pI &= \cOsqr\rhoI,
 \\
 \eqlab{cont_first}
 -\ii\omega p_1 &= -\rho_0c_0^2\div \vvv_1,
 \\
 \eqlab{navier_first}
 -\ii \omega\rho_0\vvv_1 &=
 -\grad [p_1-(\etaBO\!+\!\frac13 \etaO)\div\vvv_1]+\etaO\lap \vvv_1,
 \esubal
We make a standard Helmholtz decomposition of the velocity field $\vvv_1$,\citep{Nyborg1958, Lee1989, Bradley1996, Karlsen2015}
 \beq{Helmholtz_decomp}
 \vvv_1=\vvvl_1+\vvvd_1, \text{ where $\rot \vvvl_1=\zerovec$ and $\div\vvvd_1 =0$},
 \eeq
and insert it in \eqref{cont_navier_first}. We assume that the equations separate in solenoidal and irrotational parts and find
 \bsubal{Navier_decomp}
 \eqlab{divvls_to_p1}
 \ii\omega\kappa_0 p_1 &=\div \vvvl_1,\\
 \eqlab{p_to_vl}
 -\ii\omega\rho_0\vvvl_1&=\div\sigmat_1^d=-(1-\ii\Gamma)\grad p_1,\\
 -\ii\omega\rho_0\vvvd_1&=\div\sigmat_1^\delta=\eta_0\lap \vvvd_1.
 \esubal
From this, we derive Helmholtz equations for the bulk fields $p_1$ and $\vvvl_1$ as well as for the boundary-layer field~$\vvvd_1$,
 \bsubal{Helmholtz_eqs}
 \eqlab{Helmholtz_p}
 \lap p_1 + \kc^2 p_1 &= 0,
 & \text{where }\; & \kc =k_0\Big(1+\ii\frac{\Gamma}{2}\Big),
 \\
 \eqlab{Helmholtz_vl}
 \lap \vvvl_1+\kc^2\vvvl_1 &=\zerovec,
 && 
 \\
 \eqlab{Helmholtz_vd}
 \lap \vvvd_1+\ks^2\vvvd_1 &=\zerovec,
 & \text{where }\; &\ks =\dfrac{1+\ii}{\delta}.
 \esubal
Here, we have introduced the compressional wavenumber $\kc$ in terms of $k_0=\omega/c_0$ and $\Gamma$ defined in \eqref{Gamma_beta}, and the shear wave number $\ks$ in terms of $\delta$. Note that $\Gamma$ is of second order in $\epsilon$,
 \beq{Gamma_eps2}
 \Gamma = \frac{1+\beta}{2}\big(k_0\delta\big)^2 \sim \eps^2\ll 1.
 \eeq

From \eqref{p_to_vl} follows that the long-range velocity $\vvvl_1$ is a potential flow proportional to $\grad p_1$, and as such it is the fluid velocity of pressure acoustics. The short-range velocity $\vvvd_1$ is confined to the thin boundary layer of width $\delta$ close to the surface, and therefore it is typically not observed in experiments and is ignored in classical pressure acoustics. In the following we derive an analytic solution for the boundary-layer field $\vvvd_1$, which is used to determine a boundary condition for $p_1$. In this way, the viscous effects from the boundary layer are taken into account in computations of the long-range pressure-acoustic fields $\pI$ and $\vvvl_1$.

\subsection{Analytical form of the first-order boundary-layer field}
Using \eqref{curvi_lap_delta}, we derive an analytical solution to \eqref{Helmholtz_vd} and find that it describes a shear wave heavily damped over a single wave length, as it travels away from the surface with speed $c_\mathrm{s}=\omega\delta \ll c_0$,
 \beq{vd_sol}
 \vvvd_1=\vvvds_1 \ee^{\ii\ks\zs} + \ord{\eps}.
 \eeq
To satisfy the boundary condition \eqnoref{bc_first}, we impose the following condition for $\vvvds_1$ at the equilibrium position $\rrr=\sss_0$ of the wall,
 \beq{vvvds_sol}
 \vvvds_1 = \vvvwall_1-\vvvls_1, \text{ first-order no-slip condition}.
 \eeq

\subsection{Boundary condition for the first-order pressure field}
We now derive a boundary condition for the first-order pressure field $\pI$, which takes the viscous boundary layer effects into account without explicit reference to $\vvvI$.
First, it is important to note that the incompressibility condition $\div\vvvd_1=0$ used on \eqref{vd_sol} leads to a small perpendicular short-range velocity,
 \beq{vdsperp}
 \vdsperp = \dfrac{\ii}{\ks}\div \vvvds_1=\dfrac{\ii}{\ks}\div \vvvwall_1-\dfrac{\ii}{\ks}\div \vvvls_1.
 \eeq
In the following, we repeatedly exploit the smallness of this velocity component, $|\vdsperp|\sim \eps |v_1| \ll |v_1|$. Using the no-slip condition \eqnoref{vvvds_sol}, the boundary condition on the long-range velocity becomes,
 \bsubal{vlsperp}
 \vlsperp &=\vwallperp-\vdsperp
 \\\eqlab{vlsperp2}
 &=\Big(\vwallperp-\frac{\ii}{\ks}\div \vvvwall_1\Big)+\frac{\ii}{\ks}\div \vvvls_1
 \\\eqlab{vlsperp3}
 &\approx \Big(\vwallperp-\frac{\ii}{\ks}\pardiv \vvvwallpar\Big)+\frac{\ii}{\ks}\pardiv \vvvlspar,
 \esubal
where the last step is written for later convenience using $\frac{\ii}{\ks}\div\big(\vvvls_1-\vvvwall_1\big)
= \frac{\ii}{\ks}\pardiv\big(\vvvlspar-\vvvwallpar\big) - \frac{\ii\H_\perp}{\ks}\vdsperp$ from \eqsref{curvi_derivatives_zero_a}{vvvds_sol}. Note that this boundary condition involves the usual expression $\vwallperp$ used in classical pressure acoustics plus an $\ord{\eps}$-correction term proportional to $\ks^{-1}$, due to the parallel divergence of fluid velocity inside the boundary layer that forces a fluid flow perpendicular to the surface to fulfil the incompressibility of the short-range velocity component $\vvvd_1$. Note also that this correction term is generated partly by the external wall motion $-\frac{\ii}{\ks}\pardiv \vvvwallpar$ and partly by the fluid motion itself $\frac{\ii}{\ks}\pargrad\cdot\vvvlspar$. Hence, the wall can affect the long-range fields either by a perpendicular component $\vwallperp$ or by a parallel divergence $\pardiv \vvvwallpar$. The correction term $\frac{\ii}{\ks}\pardiv \vvvlspar$ due to the fluid motion itself gives the boundary-layer damping of the acoustic energy, see \secref{Acoustic_power_loss}.

Finally, we write \eqref{vlsperp2} in terms of the pressure $p_1$ using $\div \vvvls_1=\div \vvvl_1-\pp_\perp \vlperp$ and \eqref{Navier_decomp},
 \bal\eqlab{p1_bc}
 \pp_\perp{p_1} &= \frac{\ii\omega\rho_0}{1-\ii\Gamma} \Big(\vwallperp-\dfrac{\ii}{\ks}\div\vvvwall_1\Big)-\dfrac{\ii}{\ks}\Big(\kc^2p_1+\pp_\perp^2p_1\Big),\nn
 \\
 &\text{boundary condition at $\rrr=\sss_0\in\pp\Omega$}.
 \eal

\subsection{Boundary condition for the first-order normal stress}
The boundary condition for the first-order normal stress $\sigmat_1\cdot\ezs$ on the surrounding wall is found using \eqsref{sigma}{tau}. Here, the divergence term can be neglected, because \eqref{divvls_to_p1} leads to $|\etaO \div \vvvl_1| \approx  \frac{\etaO k_0^2}{\omega\rhoO}\:\pI \approx \Gamma \pI \ll p_1$. Further, the viscous stress is dominated by the term with $\pp_\perp\vvvd_{1}$, and we obtain
 \beq{stress1_p1_eta}
 \sigmat_1\cdot \ezs = -p_1\ezs + \etaO\pp_\perp \vvvd_{1},
 \text{ at $\rrr=\sss_0\in\pp \Omega$}.
 \eeq
Using solution~\eqnoref{vd_sol} for the short-range velocity $\vvvd_1$, we find $\pp_\perp\vvvd_{1} = \ii\ks\vvvd_{1}$, which after using \eqsref{p_to_vl}{vvvds_sol} can be expressed only with reference to the long-range pressure $p_1$ and wall velocity $\vvvwall_1$,
 \bal\eqlab{stress1_bc}
 \sigmat_1\cdot\ezs &= -p_1\ezs+
 \ii\ks \eta_0 \Big(\vvvwall_1+\dfrac{\ii}{\omega\rho_0}\grad p_1\Big),\nn
 \\
 &\text{boundary condition at $\rrr=\sss_0\in\pp \Omega$}.
 \eal
This is the usual pressure condition plus a correction term of due to viscous stress from the boundary layer.

Equations~\eqnoref{Helmholtz_eqs}, \eqnoref{vlsperp}, \eqnoref{p1_bc}, and \eqnoref{stress1_bc} constitute our main theoretical result for the first-order acoustic fields.

\section{Acoustic power loss}
\seclab{Acoustic_power_loss}
From the pressure $\pI$, we derive an expression for the acoustic power loss solely in terms of long-range fields.
First, we introduce the energy density $\Eac^d$ and the energy-flux density $\SSSac$ of the long-range acoustic fields,
 \bsubal{E(t)}
 \Eac^d(\rrr,t)&=\dfrac{1}{2} \big[\Re(p_1\eiot)\big]^2
 + \dfrac{1}{2}\rho_0 \big|\Re(\vvvl_1\eiot)\big|^2,\\
 \SSSac(\rrr,t) &=\Re\big(p_1\eiot\big)\:\Re\big(\vvvl_1\eiot\big),
 \esubal
with the time averages
 \bsubal{avrEavrS}
 \eqlab{Eacd}
 \avr{\Eac^d}&=\frac{1}{4}\kappa_0 |p_1|^2 + \frac{1}{4}\rho_0|\vvvl_1|^2,
 \\
 \avr{\SSSac} &= \avr{p_1\vvvl_1} = \cOsqr\avr{\rho_1\vvvl_1}.
 \esubal
In terms of real-valued physical quantities, \eqsref{cont_first}{p_to_vl} become $\ppt\Re(p_1\eiot)=-\rho_0\cOsqr\div \Re (\vvvl_1\eiot)$ and $\rho_0 \ppt \Re\big(\vvvl_1\eiot\big)=-\grad\Re\big[ (1-\ii\Gamma)p_1\eiot\big]$. Taking the scalar product of $\Re (\vvvl_1\eiot)$ with the latter leads to expressions for the time derivative $\ppt\Eac^d$ and its time-averaged value $\avr{\ppt \Eac^d}$, which is zero due to the harmonic time dependence,
 \bsubal{pptE_t_avr}
 \eqlab{pptEa}
 \ppt\Eac^d & = -\div \SSSac -\Gamma \rho_0\omega\big|\Re(\vvvl_1\eiot)\big|^2,
 \\
 \eqlab{avrpptE_a}
 -\div{\avr{\SSSac}} & = \frac{1}{2}\Gamma\omega\rho_0 \big|\vvvl_1\big|^2.
 \esubal
The latter expression describes the local balance between the convergence of energy flux due to pressure and the rate of change of acoustic energy due to the combined effect of viscous dissipation and viscous energy flux, see \appref{Full_E_balance} for a more detailed discussion of this point. Integrating \eqref{avrpptE_a} over the entire fluid domain $\Omega$, and using Gauss's theorem with the $\perp$-direction pointing into $\Omega$, leads to the global balance of energy rates,
 \beq{global_energy_rate}
 \int_{\pp\Omega} \avr{p_1\vlsperp} \ \dm A=\int_\Omega \frac{1}{2}\Gamma\rho_0\omega |\vvvl_1|^2\ \dm V.
 \eeq
Note that this general result only reduces to that of classical pressure acoustics in the special case where $\vlsperp = \vwallperp$. As seen from \eqref{vlsperp3}, $\vlsperp$ is generated partly externally by the wall motion, and partly internally by the fluid motion. Inserting \eqref{vlsperp3} into \eqref{global_energy_rate}, and separating wall-velocity terms from fluid-velocity terms gives,
 \bal\eqlab{PworkPloss}
 \oint_{\pp\Omega} &\Big\langle p_1\Big(\vwallperp-\dfrac{\ii}{\ks}\pardiv\vvvwallpar\Big)\Big\rangle \ \dm A
  \\ \nn
 &= \int_\Omega \frac{1}{2}\Gamma\rho_0\omega |\vvvl_1|^2\ \dm V
 -\oint_{\pp\Omega} \Big\langle p_1\Big(\dfrac{\ii}{\ks}\pardiv\vvvlspar\Big)\Big\rangle \ \dm A.
 \eal
Here, the left-hand side represents the acoustic power gain due to the wall motion, while the right-hand side represents the acoustic power loss $\avr{\Ploss^d}$ due to the fluid motion. Integrating the last term by parts and using that $\oint_{\pp\Omega} \pardiv\big\langle p_1\big(\frac{\ii}{\ks}\vvvlspar\big)\big\rangle \ \dm A=0$ for any closed surface, we can by \eqref{p_to_vl} rewrite $\avr{\Ploss^d}$ to lowest order in $\Gamma$ as,
 \beq{PlossSuper}
 \avr{\Ploss^d}=\omega\int_\Omega\frac{\Gamma}{2}\rho_0|\vvvl_1|^2 \ \dm V+\omega\oint_{\pp\Omega} \frac{\delta}{4}\rho_0 \big|\vvvlspar\big|^2\: \dm A,
 \eeq
which is always positive. The quality factor $Q$ of an acoustic cavity resonator can be calculated from the long-range fields $\avr{\Eac^d}$ in \eqref{Eacd} and $\avr{\Ploss^d}$ in \eqref{PlossSuper} as
 \beq{Q-factor}
 Q=\omegares \frac{\int_\Omega \avr{\Eac^d}\ \dm V}{\avr{\Ploss^d}}.
 \eeq
We emphasize that in general, $\avr{\Ploss}$ is not identical to the viscous heat generation $\avr{\Pdiss} = \int_\Omega \avr{\grad\vvv_1:\taut_1}\, \dm V$, although as discussed in \appref{Full_E_balance}, these might be approximately equal in many common situations\cite{Hahn2015}.

\section{Second-order streaming fields}
\seclab{second_theory}
The acoustic streaming is governed by the time-averaged part of \eqref{cont_navier} to second order in $\text{Ma} = \frac{1}{\cO}|\vvvI|$, together with the boundary condition \eqref{bc_second},
 \bsuba{cont_navier_second}
 \balat{2}{
 0		&=\div\big(\rho_0\vvv_2+\avr{\rho_1\vvv_1}\big),
 &&\text{for $\rrr \in \Omega$},
 \\
 \zerovec&=  \div\sigmat_2-\rho_0\div\avr{\vvv_1\vvv_1},\!
 &\quad&\text{for $\rrr \in \Omega$},
 \\
 \eqlab{Langrage_bc_v2}
 \zerovec&=\vvv_2+\avr{(\sss_1\cdot\grad)\vvv_1},
 &&\text{for $\rrr = \sss_0 \in \pp\Omega$}.
 }\ealat
 \esuba
Again, we make a decomposition into long-range bulk fields "$d$" and  short-range boundary-layer fields "$\delta$",
 \bsubal{decomp_second}
 \vvv_2&=\vvvl_2+\vvvd_2,\\
 p_2&= \pl_2+\pd_2,\\
 \sigmat_2&=\sigmatl_2+\sigmatd_2,
 \\
 \eqlab{decomp_bc}
 \vvvls_2 &= -\vvvds_2 - \avr{(\sss_1\cdot\grad)\vvv_1},
 \text{ at $\rrr = \sss_0 \in \pp\Omega$},
 \esubal
but in contrast to the first-order decomposition~\eqnoref{Helmholtz_decomp}, the second-order length-scale decomposition~\eqnoref{decomp_second} is not a Helmholtz decomposition. Nevertheless, the computational strategy remains the same: we find an analytical solution to the short-range "$\delta$"-fields, and from this derive boundary conditions on the long-range "$d$"-fields.

Note that our method to calculate the steady second-order fields differs from the standard method of matching "inner" boundary-layer solutions with "outer" bulk solutions. Our short- and long-range fields co-exist in the boundary layer, but are related by imposing boundary conditions on the instantaneous fluid-solid interface.

\subsection{Short-range boundary-layer streaming}
The short-range part of \eqref{cont_navier_second} consists of all terms containing at least one short-range "$\delta$"-field,
 \bsub
 \eqlab{cont_navier_second_short}
 \bal
 \eqlab{cont_navier_second_short_a}
 0		&=\div(\rho_0\vvvd_2+\avr{\rho_1\vvvd_1}),
 \\
 \eqlab{cont_navier_second_short_b}
 \zerovec &= -\rho_0\div\avr{\vvvd_1\vvvd_1+\vvvd_1\vvvl_1+\vvvl_1\vvvd_1} + \div\sigmatd_2,
 \\
 \eqlab{sigmad2}
 \div \sigmatd_2 &= \nablabf\big(-\pd_2+\beta\eta_0\div\vvvd_2\big)
 + \etaO \Lapl \vvvd_2,
 \\
 \eqlab{cont_navier_second_short_c}
 & \text{where $\vvvd_2\rightarrow \zerovec$ as $\zs \rightarrow \infty$}.
 \eal
 \esub
Notably, condition~\eqnoref{cont_navier_second_short_c} leads to a nonzero short-range streaming velocity $\vvvds_2$ at the wall, which, due to the full velocity boundary condition~\eqnoref{Langrage_bc_v2}, in turn implies a slip condition~\eqnoref{decomp_bc} on the long-range streaming velocity $ \vvvls_2$.

First, we investigate the scaling of $\pd_2$ by taking the divergence of \eqref{cont_navier_second_short_b} and using \eqref{cont_navier_second_short_a} together with $\div \vvvd_1=0$ and \eqref{Navier_decomp},
 \bsubal{scale_p2d}
 \lap \pd_2=&-\nuO(1+\beta)\lap\avr{\vvvd_1\scap\grad\rho_1} \nonumber  \\
 \eqlab{pd2}
 &-\rho_0\div(\div\avr{\vvvd_1\vvvd_1+\vvvd_1\vvvl_1+\vvvl_1\vvvd_1})\\
 =&-\rho_0\Gamma\lap\avr{\vvvd_1\scap(\ii\vvvl_1)}
 +2\rho_0k_0^2\avr{\vvvd_1\scap\vvvl_1}\nn \\
 &-\rho_0\avr{\grad(2\vvvl_1+\vvvd_1)\scapmat(\grad\vvvd_1)^\mr{T}}.
 \esubal
Recalling that $|\vdsperp|\sim  \delta d^{-1} v_1$ from \eqref{vdsperp}, we find $|\rho_0(\grad\vvvl_1)\scapmat(\grad\vvvd_1)^\mr{T}|\sim (\delta d)^{-1}\rho_0 v_1^2$ which is the largest possible scaling of the right-hand side. Since by definition $\pd_2$ is a boundary-layer field, we have $|\lap \pd_2|\sim \delta^{-2}\pd_2$, and the maximal scaling of $|\pd_2|$ becomes,
 \bal
 |\pd_2|\lesssim \epsilon\rho_0 v_1^2.
 \eal
Thus, $\grad \pd_2$ can be neglected in the parallel component of \eqref{cont_navier_second_short_b}, but not necessarily in the perpendicular one. Similarly in \eqref{sigmad2} we have $\nablabf\big(\beta\eta_0\div\vvvd_2\big)=-\beta\nu_0\grad\avr{\vvvd_1\scap\grad\rho_1}$ which scales as $\beta\eta_0 d^{-2}\frac{v_1^2}{c_0}$ which is much smaller than $|\eta_0\lap \vvvdspar|\sim \eta_0\delta^{-2} \frac{v_1^2}{c_0}$.

Henceforth, using the approximation \eqnoref{curvi_lap_delta} for the boundary-layer field $\vvvd_{2}$ in \eqref{cont_navier_second_short_b}, we get the parallel equation to lowest order in $\eps$,
 \bsub
 \bal\eqlab{cont_navier_second_short_par}
 \nu_0 \pp_\zs^2 \vvvd_{2\parallel}=\Big[\div\avr{\vvvd_1\vvvl_1
 +\vvvl_1\vvvd_1+\vvvd_1\vvvd_1}\Big]^{{}}_\parallel.
 \eal
Combining this with \eqref{cont_navier_second_short_a}, and using \eqsref{curvi_div_delta}{Helmholtz_decomp}, leads to an equation for the perpendicular component $\vd_{2\perp}$  of the short-range streaming velocity,
 \beq{cont_navier_second_short_perp}
 \pp_\zs \vd_{2\perp} =-\pardiv\vvvd_{2\parallel}-\dfrac{1}{\rho_0}\avr{\vvvd_1\cdot\grad\rho_1}.
 \eeq
 \esub

To determine the analytical solution for $\vvvd_{2\parallel}$ in \eqref{cont_navier_second_short_par}, we need to evaluate divergence terms of the form $\div\avr{\vvvI^\alpha\vvvI^\beta}$, with $\alpha,\beta = d,\delta$. To this end, we Taylor-expand $\vvvl_1$ to first order in $\zeta$ in the boundary layer, and use the solution~\eqnoref{vd_sol} for $\vvvd_1$,
 \bsubalat{2}{
 \eqlab{vl1BL}
 \vvvl_1 &= \vvvls_1 + \big(\pp_\perp\vvvl_1\big)^0\:\zs, &\quad & \text { for } \zs \ll d,
 \\
 \eqlab{vd1BL}
 \vvvd_1 & =\vvvds_1\:q(\zeta), &\quad & \text{ with } q(\zeta) = \ee^{\ii\ks\zs}.
 }\esubalat
With these expressions, \eqref{cont_navier_second_short_par} becomes,
 \bal
 \eqlab{cont_navier_second_short_par_Taylor}
 &\nu_0 \pp_\zs^2 \vvvd_{2\parallel}
 =\Big\{\div\Big\langle
 \big[\vvvds_1q\big]\big[\vvvls_11\big]
 +\big[\vvvds_1q\big]\big[(\pp_\perp\vvvl_1)^0\zeta\big]
 \\
 \nn
 &
 +\big[\vvvls_11\big]\big[\vvvds_1q\big]
 +\big[(\pp_\perp\vvvl_1)^0\zeta\big]\big[\vvvds_1q\big]
  +\big[\vvvds_1q\big]\big[\vvvds_1q\big]\Big\rangle
 \Big\}^{{}}_\parallel.
 \eal
In general, the divergence $\div\avr{\AAA_1\BBB_1}$ of the time-averaged outer product of two first-order fields of the form $\AAA_1 = \AAAs_1(\xs,\ys)\: a(\zs)$ and $\BBB_1 = \BBBs_1(\xs,\ys)\: b(\zs)$, is
\bsubal{div_AB}
 &\div \avr{[\AAAs_1 a] [\BBBs_1 b]}
  =\dfrac{1}{2} \Re \Big\{\div \Big[\big(\AAAs_1 a\big)\big(\BBBs_1 b\big)^* \Big]\Big\}\\
 &=\dfrac{1}{2} \Re \Big\{\div \Big[(ab^*) \big(\AAAs_1 \BBB_1^{0*}\big)\Big]\Big\}\\
 &=\dfrac{1}{2} \Re \Big\{ab^* \div \big(\AAAs_1 \BBB_1^{0*} \big) +
                     \AAAs_1 \big(\BBB_1^{0*}\scap\grad\big) (a b^*)\Big\} \\
 &=\dfrac{1}{2} \Re \Big\{ab^* \div \big(\AAAs_1 \BBB_1^{0*}\big) +
                     \AAAs_1 B_{1\perp}^{0*}\pp_\perp(a b^*)\Big\}.
 \esubal
When solving for $\vvvds_{2\parallel}$ in \eqref{cont_navier_second_short_par}, we must integrate such divergences twice and then evaluate the result at the surface $\zs=0$. Straightforward integration yields
 \bsub
 \eqlab{IabnEqs}
 \bal
 \eqlab{zetaIntegrals}
 \int^{\zeta}\!\dm\zs_2 & \int^{\zeta_2}\!\dm\zs_1\:
 \div \Big[\big(\AAAs_1 a(\zs_1)\big)\big(\BBBs_1 b(\zs_1)\big)^* \Big]\bigg|^{{}}_{\zs=0}
 \nn \\
 &=\dfrac{1}{2} \Re \Big\{\intII ab  \div \big(\AAAs_1 \BBB_1^{0*}\big)
 + \intI ab\AAAs_1 B_{1\perp}^{0*}\Big\},
 \eal
where we have defined the integrals $\intn ab$ as,
 \bal
 \eqlab{def_intI}
 \intI ab &=  \int^{\zeta}\!\dm\zs_1\:
 a(\zs_1)\: b(\zs_1)^* \bigg|^{{}}_{\zs=0},
 \\
 \eqlab{def_intII}
 \intII ab &=  \int^{\zeta}\!\dm\zs_2 \int^{\zeta_2}\!\dm\zs_1\:
 a(\zs_1)\: b(\zs_1)^* \bigg|^{{}}_{\zs=0},
 \\
 \eqlab{def_intIII}
 \intIII ab &=  \int^{\zeta}\!\dm\zs_3 \int^{\zeta_3}\!\dm\zs_2 \int^{\zeta_2}\!\dm\zs_1\:
 a(\zs_1)\: b(\zs_1)^* \bigg|^{{}}_{\zs=0}.
 \eal
We choose  all integration constants to be zero to fulfil the condition \eqnoref{cont_navier_second_short_c} at infinity. From \eqref{cont_navier_second_short_par_Taylor} we see that the functions $a(\zeta)$ and $b(\zeta)$ in our case are $q(\zeta)$, $\zeta$ or unity. By straightforward integration, we find in increasing order of $\delta$,
 \balat{3}{
 \eqlab{Iabn}
  \intI qq &= -\frac12 \delta,
 \quad & \intI q1 &= -\frac{1+\ii}{2} \delta,
 \\
 \nn
 \intII qq &= \frac14 \delta^2,
 \quad & \intII q1 &= \frac{\ii}{2} \delta^2,
 \quad & \intI q\zs &= -\frac{\ii}{2} \delta^2,
 \\
 \nn
 \intIII qq &= -\frac18 \delta^3,
 \quad & \intIII q1 &= \frac{1-\ii}{4} \delta^3,
 \quad & \intII q\zs &= -\frac{1-\ii}{2} \delta^3.
 }\ealat
 \esub
Using \eqref{IabnEqs} and $\vdsperp \sim \eps\big|\vvvdspar|$ from
\eqref{vdsperp}, we find $\vvvds_{2\parallel}$ by integration of \eqref{cont_navier_second_short_par_Taylor} to leading order in $\eps$,
 \bal
 \vvvds_{2\parallel} = \;&
 \frac{1}{2\nu_0}\Re \Big\{
 \intII qq  \div\big(\vvvds_1 \vvvdsC_1\big)+
 \intII q1 \div\big(\vvvds_1 \vvvlsC_1\big)
 \nn
 \\
 &+ \intII 1q \div\big(\vvvls_1 \vvvdsC_1\big)
 +\intI qq \vvvds_1\vdsCperp
 + \intI 1q \vvvls_1\vdsCperp
 \nn
 \\
 &+ \intI q1 \vvvds_1\vlsCperp
 + \intI q\zs \vvvds_1 \pp_\perp \vlCperp
\Big\}_\parallel.
 \eal
Remarkably, the term $\intI q1 \vvvds_{1}\vlsCperp$ scales with a factor $\eps^{-1}$ compared to all other terms, and thus may dominate the boundary-layer velocity. However, in the computation of the long-range slip velocity $\vvvls_{2\parallel}$ in \secref{LongRangeStreaming}, its contribution is canceled by the Stokes drift $\avr{\sss_1\cdot\grad\vvv_1}$, as also noted  in Ref.~[\onlinecite{Vanneste2011}]. Using $\vvvls_1=\vvvwall_1-\vvvds_1$, the property $(\intn ab)^*=\intn ba$, and rearranging terms gives,
 \bal
 \eqlab{eval_par_streaming}
 \vvvds_{2\parallel} =
 \frac{1}{2\nu_0} & \Re \Big\{
 \Big(\intII qq -2\Re \intII q1\Big)  \div\big(\vvvds_1 \vvvdsC_1\big)
 \nn
 \\
 &+ \intII q1 \div\big(\vvvds_1 \vvvwallC_1\big)
 + \intII 1q \div\big(\vvvwall_1 \vvvdsC_1 \big)
 \nn
 \\
 &+\Big(\intI qq -2\Re \intI q1\Big)  \vvvds_1\vdsCperp
 + \intI 1q \vvvwall_1\vdsCperp
 \nn
 \\
 &+ \intI q1 \vvvds_1\vwallC_{1\perp}
 + \intI q\zs \vvvds_1 \pp_\perp \vlCperp
 \Big\}_\parallel.
 \eal

The perpendicular short-range velocity component $\vds_{2\perp}$ is found by integrating \eqref{cont_navier_second_short_perp} with respect to $\zeta$. The integration of the $\pardiv\vvvd_{2\parallel}$-term is carried out by simply increasing the superscript of the $I_{ab}^{(n)}$-integrals in \eqref{eval_par_streaming} from "$(n)$" to "$(n+1)$", while the integration of the $\grad\rho_1$-term is carried out by using \eqref{p_to_vl} to substitute $\frac{1}{\rhoO}\:\grad\rhoI$ by $\ii\omega c_0^{-2}\:\vvvl_1$ and introducing the suitable $I_{ab}^{(n)}$-integral for the factor $q(\zeta)\:\ii$, namely $\intI q\ii =-\ii\intI q1$,
 \bal
 \eqlab{eval_perp_streaming}
 \vds_{2\perp} =\; &
 -\frac{1}{2\nu_0} \pardiv\Re \Big\{
 \Big(\intIII qq -2\Re \intIII q1\Big)  \div\big(\vvvds_1 \vvvdsC_1\big)
 \nn
 \\
 &\qquad\qquad + \intIII q1 \div\big(\vvvds_1 \vvvwallC_1\big)
 + \intIII 1q \div\big(\vvvwall_1 \vvvdsC_1 \big)
 \nn
 \\
 &\qquad\qquad +\Big(\intII qq -2\Re \intII q1\Big)  \vvvds_1\vdsCperp
 + \intII 1q \vvvwall_1\vdsCperp
 \nn
 \\
 &\qquad\qquad + \intII q1 \vvvds_1\vwallC_{1\perp}
 + \intII q\zs \vvvds_1 \pp_\perp \vlCperp
 \Big\}_\parallel
 \nn
 \\
 &+\frac{k_0}{2\cO} \Re \Big\{ \ii\intI  q1 \vvvds_1\cdot\vvvlsC_1\Big\}.
 \eal
Evaluation of the expressions~\eqsnoref{eval_par_streaming}{eval_perp_streaming}  for $\vvvds_{2\parallel}$ and  $\vds_{2\perp}$ is straightforward. Using \eqref{Iabn}, the analytical expressions for the short-range streaming at the surface $\zs=0$ become,
 \bsub
 \eqlab{short_range_bc}
 \bal
 \vvvds_{2\parallel} =
 \frac{1}{2\omega} & \Re \Big\{
 \frac12\div\big(\vvvds_1 \vvvdsC_1\big)
 + \ii\div\big(\vvvds_1 \vvvwallC_1\big)
 \nn
 \\
 &
 - \ii\div\big(\vvvwall_1 \vvvdsC_1 \big)
 +\frac{1}{\delta}  \vvvds_1\vdsCperp
 -\ii \vvvds_1 \pp_\perp \vlCperp
 \nn
 \\
 &
 -\frac{1-\ii}{\delta}\vvvwall_1\vdsCperp
 -\frac{1+\ii}{\delta} \vvvds_1\vwallC_{1\perp}
 \Big\}_\parallel,
 \eal
and
 \bal
 \eqlab{short_range_bc_b}
 \vds_{2\perp} = &
 -\frac{\delta}{2\omega} \Re\Bigg[\!\pardiv \bigg\{\!
 -\frac{5}{4} \div\!\big(\vvvds_1 \vvvdsC_1\big)
 \\
 & \quad
 + \frac{1\!-\!\ii}{2}\div\!\big(\vvvds_1 \vvvwallC_1\big)
 + \frac{1\!+\!\ii}{2}\div\!\big(\vvvwall_1 \vvvdsC_1 \big)
 \nn
 \\
 & \quad
 + \frac{1}{2\delta}\vvvds_1\vdsCperp
 -\frac{\ii}{\delta} \vvvwall_1\vdsCperp
 + \frac{\ii}{\delta} \vvvds_1\vwallC_{1\perp}
 \nn
 \\
 & \quad
 -(1\!-\!\ii)\vvvds_1 \pp_\perp \vlCperp \bigg\}_\parallel
 -k_0^2(1\!-\!\ii) \vvvds_1\!\scap\vvvlsC_1\Bigg]
 \nn
 \\
 &=-\frac{1}{2\omega} \Re\Big[\pardiv\big(\ii\vvvds_{1\parallel}\vwallC_{1\perp}\big)\Big] +\ord{\eps}.
 \eal
\esub

\subsection{Long-range bulk streaming}
\seclab{LongRangeStreaming}
The long-range part of \eqref{cont_navier_second} is,
 \bsubal{cont_navier_second_long}
 \eqlab{cont_navier_second_long_a}
 0		&=\div[\rho_0\vvvl_2+\avr{\rho_1\vvvl_1}],
 \\
 \eqlab{cont_navier_second_long_b}
 \zerovec&= -\rho_0\div\!\avr{\vvvl_1\vvvl_1}+ \div\sigmatl_2,
 \\
 \eqlab{divsigmal2}
 \div \sigmatl_2 &= -\nablabf\big(\pl_2-\beta\eta_0\div\vvvl_2\big)
 + \etaO \Lapl \vvvl_2,
 \\
 \eqlab{cont_navier_second_long_c}
 \vvvls_2&=-\vvvds_2-\avr{(\sss_1\cdot\grad)\vvv_1},
 \text{ at $\rrr = \sss_0 \in\pp \Omega$}.
 \esubal
In contrast to the limiting-velocity  matching at the edge of the boundary layer done by Nyborg~\citep{Nyborg1958}, we define the boundary condition~\eqnoref{cont_navier_second_long_c} on the long-range streaming  $\vvvl_2$ at the equilibrium position $\rrr=\sss_0$.

We first investigate the products of first-order fields in \eqref{cont_navier_second_long}. Using \eqref{avrpptE_a} in \eqref{cont_navier_second_long_a}, we find
 \beq{div_vvvl2}
 \div\vvvl_2 = -\frac{\div\avr{\rho_1\vvvl_1}}{\rhoO} =
 -\frac{\div{\avr{\SSSac}}}{\rho_0\cO^2}=\Gamma\:\frac{k_0|\vvvl_1|^2}{2c_0}.
 \eeq
Since each term in $\div{\vvvl_2}$ scales as $ \frac{k_0}{c_0}|\vvvl_1|^2 \gg \frac{\Gamma}{2}\:\frac{k_0}{c_0}|\vvvl_1|^2$, we conclude that $\div{\vvvl_2} \approx 0$ is a good approximation corresponding to ignoring the small viscous dissipation in the energy balance expressed by \eqref{avrpptE_a}. A similar scaling leads to $\beta\eta_0\grad(\div \vvvd_2)\ll \eta_0\lap \vvvl_2$ so $\beta\eta_0\grad(\div \vvvd_2)$ can be ignored in \eqref{divsigmal2}. Finally, the divergence of momentum flux in \eqref{cont_navier_second_long_b} can be rewritten using~\eqref{p_to_vl},
 \bal\eqlab{calc_rhovv}
 \rho_0\div\avr{\vvvl_1\vvvl_1}&= -\grad \avr{\mathcal{L}^d_\mathrm{ac}}-\frac{\Gamma\omega}{\cO^2}\avr{\SSSac},
 \eal
where we introduced the long-range time-averaged acoustic Lagrangian,
\bal
\avr{\Lac^d}&=\frac{1}{4}\kappa_0 |p_1|^2 - \frac{1}{4}\rho_0|\vvvl_1|^2.
\eal
Note that $|\grad \avr{\mathcal{L}^d_\mathrm{ac}}|\sim \frac{\omega p_1^2}{\rho_0c_0^3}$ whereas $|\frac{\Gamma\omega}{\cO^2}\avr{\SSSac}|\sim \Gamma \frac{\omega p_1^2}{\rho_0c_0^3}$, so the first term in \eqref{calc_rhovv} is much larger than the second term. However, as also noted by Riaud \etal\cite{Riaud2017a}, since the first term is a gradient, it is simply balanced hydrostatically by the second order long-range pressure $\pl_2$ and therefore it can not drive any streaming velocity. In practice, it is therefore advantageous to work with the excess pressure $\pl_2-\avr{\mathcal{L}^d_\mathrm{ac}}$. With these considerations, Eqs. \eqnoref{cont_navier_second_long} become those of an incompressible Stokes flow driven by the body force $\frac{\Gamma\omega}{\cO^2}\avr{\SSSac}$ and the velocity boundary condition,
 \bsubal{cont_navier_second_long_simple}
 \eqlab{cont_navier_second_long_simple_a}
 0 &= \div\vvvl_2,
 \\
 \eqlab{cont_navier_second_long_simple_b}
 \zerovec &= -\grad\Big[\pl_2-\avr{\mathcal{L}^d_\mathrm{ac}}\Big] + \etaO\Lapl\vvvl_2
 +\frac{\Gamma\omega}{\cO^2}\avr{\SSSac},
 \\
 \eqlab{cont_navier_second_long_simple_c}
 \vvvls_2&=-\vvvds_2-\avr{(\sss_1\cdot\grad)\vvv_1}\big|^{{}}_{\zeta=0}.
 \esubal
These equations describe acoustic streaming in general. The classical Eckart streaming~\citep{Eckart1948} originates from the body force $\frac{\Gamma\omega}{\cO^2}\avr{\SSSac}$, while the classical Rayleigh streaming~\citep{LordRayleigh1884} is due to the boundary condition~\eqnoref{cont_navier_second_long_simple_c}.

The Stokes drift $\avr{\sss_1\scap\grad\vvv_1}\big|_{\zeta=0}$, induced by the oscillating wall, is computed from Eqs.~\eqnoref{dtsss}, \eqnoref{Helmholtz_decomp}, and \eqnoref{vd_sol},
 \bal
 \eqlab{eval_sdivv}
 &\avr{\sss_1\scap\grad\vvv_1}\big|^{{}}_{\zeta=0}
 = \frac{-1}{2\omega}\Re\Big[\ii\vvvwallC_1\scap\grad
 \big(\vvvl_1 + \vvvds_1 q\big)\Big]^{{}}_{\zeta=0}
 \\
 \nn
 &
 \quad = -\frac{1}{2\omega}\Re\Big[
 \ii\vvvwallC_1\scap\grad \big(\vvvl_1 + \vvvds_1\big)
 - \frac{1+\ii}{\delta}\vwallperpC\vvvds_1\Big].
 \eal
From this, combined with \eqsref{short_range_bc}{cont_navier_second_long_simple_c}, follows the boundary condition $\vvvls_2$ for the long-range streaming velocity $\vvvl_2$ expressed in terms of the short-range velocity $\vvvds_2$ and the wall velocity $\vvvwall_1$. The parallel component is
 \bsub
 \eqlab{vl2_bc}
 \bal
 \eqlab{vl2par}
 \vvvls_{2\parallel} =
 -\frac{1}{2\omega} & \Re \Big\{
 \div\Big(\frac12\vvvds_1 \vvvdsC_1
 + \ii\vvvds_1 \vvvwallC_1
 - \ii\vvvwall_1 \vvvdsC_1 \Big)
 \nn
 \\
 &
 +\frac{1}{\delta}  \vvvds_1\vdsCperp
 -\ii \vvvds_1 \pp_\perp \vlCperp
 -\frac{1-\ii}{\delta}\vvvwall_1\vdsCperp
 \nn
 \\
 &
 -\ii\vvvwallC_1\scap\grad \big(\vvvl_1 + \vvvds_1\big)
 \Big\}_\parallel,
 \eal
where the large terms proportional to $\frac{1+\ii}{\delta}\vwallperpC\vvvds_{1\parallel}$ canceled out, as also noted by Vanneste and B\"{u}hler~\citep{Vanneste2011}. Similarly, the perpendicular component becomes
 \bal
 \eqlab{vl2perp1}
 & \vls_{2\perp} =
 \frac{\delta}{2\omega} \Re\Bigg[
 -k_0^2 (1\!-\!\ii) \vvvds_1\!\cdot\vvvlsC_1
 \nn
 \\
 &\;
 +\pardiv \bigg\{\div\!\Big[-\frac{5}{4} \vvvds_1 \vvvdsC_1
 + \frac{1\!+\!\ii}{2}\Big(\vvvwall_1 \vvvdsC_1 + \vvvdsC_1\vvvwall_1 \Big)\Big]
 \nn
 \\
 &
 \quad +\!\bigg[\frac{1}{2\delta}\vdsCperp
 +\frac{\ii}{\delta}\vwallC_{1\perp}
 -(1\!-\!\ii)\pp_\perp \vlCperp\bigg]\vvvds_1
 -\frac{\ii}{\delta} \vdsCperp\vvvwall_1 \bigg\}_\parallel
 \Bigg]
 \nn
 \\
 &\; +\frac{1}{2\omega}\Re\Big[
 \ii\vvvwallC_1\scap\grad \big(\vvvl_1 + \vvvds_1\big)
 - \frac{1\!+\!\ii}{\delta}\vwallperpC\vvvds_1\Big]_{\perp}
 \\
 \eqlab{vl2perp}
 &= \frac{1}{2\omega}\Re\bigg[
 \pardiv\big(\ii\vvvds_{1\parallel}\vwallC_{1\perp}\big)
 - \frac{1\!+\!\ii}{\delta}\vwallperpC\vds_{1\perp}
 \nn
 \\
 &\qquad \qquad
 +\Big\{\ii\vvvwallC_1\scap\grad \big(\vvvl_1 + \vvvds_1\big)\Big\}_{\perp}
\bigg]
 +\ord{\eps}.
 \eal
 \esub
Taking the divergences in \eqref{vl2par} and using \eqref{vdsperp}, as well as computing \eqref{vl2perp} to lowest order in $\eps$, leads to the final expression for the slip velocity,
 \bal
 \eqlab{vl2_bc_final}
 \vvvls_2 &= \big(\AAA\cdot\een_\xi\big)\:\een_\xi + \big(\AAA\cdot\een_\eta\big)\:\een_\eta
 + \big(\BBB\cdot\een_\zeta\big)\:\een_\zeta,
 \\ \nn
 \AAA &= -\frac{1}{2\omega}  \Re \bigg\{ \vvvdsC_1\scap\grad\Big(\frac{1}{2}\vvvds_1-\ii\vvvwall_1\Big)
 -\ii\vvvwallC_1 \scap\grad \vvvl_1
 \\
 \nn
 &\qquad +\bigg[\frac{2-\ii}{2}\grad\scap\vvvdsC_1
 +\ii\Big(\grad\scap\vvvwallC_1-\pp_\perp v^{d*}_{1\perp}\Big)\bigg]\vvvds_1 \bigg\},
 \\
 \nn
 \BBB &=
 \frac{1}{2\omega} \Re\bigg\{
 \ii\een_\zeta\Big(\vvvds_1\scap\grad\Big)\vwallC_{1\perp}
 +\ii\vvvwallC_1\scap\grad\Big(\vvvds_1+\vvvl_1\Big) \bigg\},
 \eal
where $\AAA$ and $\BBB$ are associated with the parallel and perpendicular components $\vvvls_{2\parallel}$ and $\vvvls_{2\perp}$, respectively,
and where we to simplify used $\big(\vvvds_{1\parallel}\scap\pargrad\big)\vwallperpC=\big(\vvvds_{1}\scap\grad\big)\vwallperpC$.

Equations \eqnoref{cont_navier_second_long_simple} and \eqnoref{vl2_bc_final} constitute our main theoretical result for the second-order acoustic streaming.

\section{Special cases}
\seclab{Special_cases}
In the following, we study some special cases of our main results~\eqsnoref{Helmholtz_p}{p1_bc} for the acoustic pressure $\pI$ and \eqsref{cont_navier_second_long_simple}{vl2_bc_final} for the streaming velocity $\vvvl_2$, and relate them to previous studies in the literature.

\subsection{Wall oscillations restricted to the perpendicular direction}
The case of a weakly curved wall oscillating only in the perpendicular direction was studied by Nyborg~\citep{Nyborg1958} and later refined by Lee and Wang~\citep{Lee1989}. Using our notation, the boundary conditions used in these studies were
 \bsubal{OscPerpBC}
 \eqlab{OscPerpBC_Vwall}
 \vvvwall_1 &= \vlsperp\:\een_\perp,\\
 \eqlab{OscPerpBC_vd}
 \vvvds_1 &= -\vvvls_{1\parallel}.
 \esubal
For $\pI$, using \eqsref{p_to_vl}{OscPerpBC_Vwall}, we obtain $\vvvwall_1 = -\frac{\ii}{\omega\rhoO} \pp_\perp\pI\:\een_\perp$ and $\div\vvvwall_1 = \H_\perp \vlsperp$, whereby our boundary condition~\eqnoref{p1_bc} to lowest order in $\Gamma$ becomes,
 \beq{p1_bc_perp_wall}
 \pp_\perp{p_1} = \ii\omega\rho_0 \vwallperp
 - \frac{1+\ii}{2}\delta\Big(
 \kc^2\pI + \H_\perp\pp_\perp\pI + \pp_\perp^2\pI\Big).
 \eeq
Similarly for the steady streaming $\vvvl_2$, we use \eqref{OscPerpBC_vd} to substitute all occurrences of $\vvvds_1$ in the boundary condition~\eqref{vl2_bc_final} by $-\vvvlspar$. Note that we then obtain $\div\vvvds_1 = -\pargrad\cdot \vvvls_{1\parallel}=-\big(\grad\cdot\vvvl_1-\pp_\perp \vlperp-\H_\perp \vwallperp\big)$ evaluated at $\zeta = 0$. Combining this expression with the derivative rule~\eqnoref{curvi_derivatives_zero_b} and the index notation $\bar{\xs} = \ys$ and $\bar{\ys} = \xs$, as well as $\alpha$, $\beta$ = $\xs$, $\ys$, the boundary condition~\eqnoref{vl2_bc_final} for the tangential components becomes,
 \bsub
 \bal\eqlab{vls2par_final_perp_wall}
 &\vls_{2\beta} = -\frac{1}{4\omega}  \Re \bigg\{
 \vlsC_{1\alpha}\big(\ppti_\alpha \vls_{1\beta} \big)
 + \vlsC_{1\alpha} \vls_{1\bar{\beta}} T_{\alpha\bar{\beta}\beta}
 \\ \nn
 &
 -2\ii\vwallC_{1\perp} \pp_\perp \vds_{1\beta}
 +(1-2\ii)  \vdsC_{1\alpha} \vwall_{1\perp} T_{\alpha\perp\beta}
 \\ \nn
 &\;+\Big[(2\!-\!\ii)\div\vvvlC_1
 - (2\!-\!3\ii)\pp_\perp\vlCperp
 - (2\!+\!\ii)\H_\perp\vwallC_{1\perp}\Big]\vl_{1\beta}\bigg\}.
 \eal
and for the perpendicular component,
 \bal\eqlab{vls2perp_final_perp_wall}
 \vls_{2\perp}= \frac{1}{2\omega} \Re\Big\{
 \ii\vlsC_{1\alpha}\ppti_\alpha\vwallperp
 +\ii\vwallC_{1\perp}\pp_\perp\vlperp \Big\}.
 \eal
 \esub

When comparing our expressions with the results of Lee and Wang~\citep{Lee1989}, denoted by a superscript "\text{LW}" below, we note the following. Neither the pressure $\pI$ nor the steady perpendicular streaming velocity $v^d_{2\perp}$ were studied by Lee and Wang, so our results \eqsref{p1_bc_perp_wall}{vls2perp_final_perp_wall} for these fields represent an extension of their work. The slip condition~\eqnoref{vls2par_final_perp_wall} for the parallel streaming velocity $\vl_{2\beta}$ with $\beta = \xi, \eta$ is presented in Eqs.~(19)$^\mr{LW}$ and (20)$^\mr{LW}$ as the limiting values $u_L$ and $v_L$ for the two parallel components of $\vvvl_2$ outside the boundary layer. A direct comparison is obtained by: (1) Identifying our $\vvvl_1$ with the acoustic velocity $(u_{a0}, v_{a0}, w_{a0})$ in LW, and our $T_{kji}$ with $T_{ijk}$ in LW; (2) Taking the complex conjugate of the argument of the real value in \eqref{vls2par_final_perp_wall}, and (3) noting that $q_x$ and $q_y$ defined in Eqs.~(3)$^\mr{LW}$ and (4)$^\mr{LW}$ equal the first two terms of \eqref{vls2par_final_perp_wall}. By inspection we find agreement, except that Lee and Wang are missing the terms $-2\ii\vwallC_{1\perp} \pp_\perp \vds_{1\beta}  +(1-2\ii)  \vdsC_{1\alpha} \vwall_{1\perp} T_{\alpha\perp\beta}$. The two terms with the prefactor "$2\ii$" arise in our calculation from the Lagrangian velocity boundary condition~\eqnoref{Langrage_bc_v2}, where Lee and Wang have used the no slip condition $\vvv_2 = \zerovec$, while the remaining term $\vdsC_{1\alpha} \vwall_{1\perp} T_{\alpha\perp\beta}$ is left out by Lee and Wang without comment.

\subsection{A flat wall oscillating in any direction}
The case of a flat wall oscillating in any direction was studied by Vanneste and B\"uhler~\citep{Vanneste2011}. In this case, we adapt Cartesian coordinates $(\xs,\ys,\zs) \rightarrow (x,y,z)$, for which all scale factors $h_i$ are unity, $\ppti_i = \pp_i$, and all Christoffel symbols $T_{kji}$ are zero. The resulting expressions for the boundary conditions~\eqnoref{p1_bc} for the pressure and~\eqnoref{p1_bc} for the long-range streaming $\vvvl_2$ then simplify to
 \bsubal{bc_flat_wall}
 \eqlab{p1_bc_flat_wall}
 \pp_\perp{p_1} &= \ii\omega\rho_0 \vwallperp
 - \frac{1+\ii}{2}\delta\big(\ii\omega\rho_0
 \div\vvvwall_1+\kc^2p_1+\pp_\perp^2p_1\big),
 \\
 \eqlab{vls2par_final_flat_wall}
 \vls_{2\beta} &= -\frac{1}{4\omega}  \Re \bigg\{
 (1-2\ii)\vdsC_{1\alpha}\pp_\alpha \vds_{1\beta}
 -4\ii \vdsC_{1\alpha}\pp_\alpha \vls_{1\beta}
 \nn
 \\
 +\Big[(2+&\ii)\pp_\alpha\vdsC_{1\alpha}
 +2\ii\big(\pp_\alpha \vlsC_{1\alpha}\!-\!\pp_\perp\!\vlCperp\big)\Big]\vds_{1\beta}
 -2\ii\:\vlC_{1k}\pp_k\vl_{1\beta}\!\bigg\},
 \\
 \eqlab{vls2perp_final_flat_wall}
 \vls_{2\perp}&= -\frac{1}{4\omega}
 \Re\Big\{-2\ii\:\vlC_{1k}\pp_k\vlperp \Big\}.
 \esubal
The pressure condition~\eqnoref{p1_bc_flat_wall} was not studied in Ref.~\onlinecite{Vanneste2011}, so it represents an extension of the existing theory. On the other hand, \eqsref{vls2par_final_flat_wall}{vls2perp_final_flat_wall} are in full agreement with Eq.~(4.14) in Vanneste and B\"uhler~\citep{Vanneste2011}. To see this, we identify our first-order symbols with those used in Ref.~\onlinecite{Vanneste2011} as $\vvvl_1 \leftrightarrow 2\nablabf \hat{\phi}$ and $\vvvds_{1\parallel} \leftrightarrow  -2\hat{U}_1\een_x -2\hat{V}_1\een_y$, and we relate our steady Eulerian second-order long-range velocity $\vvvl_2$ with their Lagrangian mean flow $\bar{\uuu}^\mr{L}$ using the Stokes drift expression~\eqnoref{Langrage_bc_v2} as $\vvvl_2 + \frac{1}{\omega} \avr{\ii\vvvl_1\cdot \nablabf \vvvl_1} \leftrightarrow  \:\bar{\uuu}^\mr{L}$ at the interface $z = 0$.

\subsection{Small surface velocity compared to the bulk velocity}
At resonance in acoustic devices with a large resonator quality factor $Q \gg 1$, the wall velocity $\vvvwall_1$ is typically a factor $Q$ smaller than the bulk fluid velocity $\vvvl_1$,\citep{Muller2013, Muller2015}
$\vwall_1 \sim Q^{-1} \vl_1 \ll \vl_1$. In this case, as well as for rigid walls, we use $\vvvwall_1 = \zerovec$ in \eqref{vl2_bc_final}, so that $\vvvds_1\approx -\vvvls_1$ and
\bal
 \eqlab{resonance_approx}
 \avr{\vvvds_1\scap\grad\vvvds_1}
 &\approx \avr{\vvvls_1\scap\grad\vvvls_1}
 =\avr{\vvvlspar\scap\pargrad \vvvlspar+\vvvlspar\scap\pargrad\vvvls_{1\perp}}
 \nn \\
 &\approx \frac{1}{4} \pargrad|\vvvlspar|^2.
\eal
Here, we have neglected $\vlsperp$ because $|\vlsperp| \approx |\vwallperp| \ll \big|\vvvlspar\big|$ and used that $\rot\vvvl_1=\zerovec$ from \eqref{Helmholtz_decomp}. Hence, the slip-velocity $\vvvls_2$ for devices with rigid walls $\vvvwall_1 = \zerovec$, or resonant devices with $|\vvvls_1|\gg|\vvvwall_1|$, becomes
 \bsub
 \eqlab{vl2_bc_V_zero2}
 \bal
 \eqlab{vls2par_V_zero2}
 \vvvls_{2\parallel} &=
 \!\!\frac{-1}{8\omega}\pargrad\big|\vvvlspar \big|^2
 \!\!-\Re\bigg\{\hspace{-1mm}\bigg(\!\frac{2-\ii}{4\omega}\pargrad\scap\vvv_{1\parallel}^{d0*}
 \!+\hspace{-0.5mm}\frac{\ii}{2\omega}\pp_\perp v^{d*}_{1\perp}\bigg)\vvvlspar\!\bigg\},
 \\
 \eqlab{vls2perp_V_zero2}
 \vls_{2\perp} &= 0.
 \eal
 \esub
Two important limits are parallel acoustics, where $|\pp_\perp v^{d}_{1\perp}|\ll |\pargrad\cdot \vvvlspar|$, and perpendicular acoustics, where $|\pp_\perp v^{d}_{1\perp}|\gg |\pargrad\cdot \vvvlspar|$. In the first limit, the pressure is mainly related to the parallel velocity variations and from \eqsref{divvls_to_p1}{p_to_vl} we have $\pargrad\cdot\vvv_{1\parallel}^d= \ii\omega\kappa_0 p_1$ and $\vvvlspar=-\frac{\ii}{\rho_0\omega}\pargrad p_1$. For parallel acoustics we can therefore write \eqref{vls2par_V_zero2} as,
\bsub
 \bal\eqlab{vlspar_parallel_simple}
 &\vvvls_{2\parallel}= \dfrac{1}{8\omega\rho_0}\pargrad \big(2\kappa_0\big|p_1\big|^2-
 \rho_0\big|\vvvlspar \big|^2 \big)+\frac{\kappa_0}{2}\avr{\SSS^d_{\mr{ac}\parallel}},
 \\ \nn
 &\text{for parallel acoustics, } |\pp_\perp v^{d}_{1\perp}|\ll |\pargrad\cdot \vvvlspar|.
 \eal
The classical period-doubled Rayleigh streaming\citep{LordRayleigh1884}, which arises from a one-dimensional parallel standing wave, results from the gradient-term in \eqref{vlspar_parallel_simple}. This is seen by considering a rigid wall in the $x$-$y$ plane with a standing wave above it in the $x$ direction of the form $\vvvl_1 = v_{1a}\cos(k_0 x)\:\een_x$, where $v_{1a}$ is a velocity amplitude. Inserting this into \eqref{vlspar_parallel_simple} yields Rayleigh's seminal boundary velocity $\vvvls_{2\parallel} = \frac38 \frac{v^2_{1a}}{\cO}\:\sin(2k_0x)\: \een_x$. Another equally simple example of parallel acoustics is the boundary condition generated by a planar travelling wave of the form $\vvvl_1 = v_{1a}\eikOx\:\een_x$. Here, only the energy-flux vector in \eqref{vlspar_parallel_simple} contributes to the streaming velocity which becomes the constant value $\vvvls_{2\parallel} = \frac14 \frac{v^2_{1a}}{\cO}\: \een_x$.\\

The opposite limit is perpendicular acoustics, where the pressure is mainly related to the perpendicular velocity variations $\pp_\perp v^{d}_{1\perp}= \ii\omega\kappa_0 p_1$. In this limit, \eqref{vls2par_V_zero2} is given by a single term,
 \bal\eqlab{vlspar_perp_simple}
 &\vvvls_{2\parallel}=-\kappa_0\avr{\SSS^d_{\mr{ac}\parallel}},
 \\ \nn
 &\text{for perpendicular acoustics, }
 |\pp_\perp v^{d}_{1\perp}|\gg |\pargrad\cdot \vvvlspar|.
 \eal
\esub
We emphasize that in these two limits, the only mechanism that can induce a streaming slip velocity, which rotates parallel to the surface, is the energy-flux-density vector $\avr{\SSS^d_{\mr{ac}}}$. As seen from \eqref{cont_navier_second_long_simple_b}, this mechanism also governs the force density driving streaming in the bulk. In general, $\avr{\SSS^d_{\mr{ac}}}$ can drive rotating streaming if it has a nonzero curl, which we calculate to lowest order in $\Gamma$ using \eqref{p_to_vl} and $\rot\vvvl_1=\zerovec$, and find to be proportional to the acoustic angular momentum density,
 \bal
 \rot\avr{\SSS^d_{\mr{ac}}} =
 \omega^2 \avr{\rrr^d_1\times (\rho_0\vvvl_1)},
 \qquad
 \rrr^d_1 =
 \frac{\ii}{\omega}\vvvl_1.
 \eal

\section{Numerical modeling in COMSOL}
\seclab{Comsol_model}

In the following we implement our extended acoustic pressure theory, \eqsref{Helmholtz_p}{p1_bc} for $p_1$, and streaming theory, \eqsref{cont_navier_second_long_simple}{vl2_bc_final} for $\vvvl_2$ and $p_2$, in the finite-element method (FEM) software COMSOL Multiphysics\cite{COMSOL53a}. We compare these simulations with a full boundary-layer-resolved model for the acoustics, \eqsref{cont_navier_first}{bc_first} for $\vvv_1$ and $p_1$, and for the streaming, \eqsref{cont_navier_second}{bc_second} for $\vvv_2$ and $p_2$, where the full model is based on our previous acoustofluidic modeling of fluids-only systems \cite{Muller2012, Muller2014, Muller2015} and solid-fluid systems \cite{Ley2016}.

Remarkably, our extended (effective) acoustic pressure model makes it possible to simulate acoustofluidic systems not accessible to the brute-force method of the full model for three reasons: (1) In the full model, the thin boundary layers need to be  resolved with a fine FEM mesh. This is not needed in our effective model. (2) For the first-order acoustics, the full model is based on the vector field $\vvv_1$ and the scalar field $p_1$, whereas our effective model is only based on the scalar field $p_1$. (3) For the second-order streaming, the full equations \eqnoref{cont_navier_second} contain large canceling terms, which have been removed in the equations \eqnoref{cont_navier_second_long_simple} used in the effective model. Therefore, also in the bulk, the effective model can  be computed on a much coarser FEM mesh than the full model.

In \secref{ExI}, we model a fluid domain $\Omgf$ driven by boundary conditions applied directly on $\pp\Omgf$, and in \secref{ExII}, we model a fluid domain $\Omgf$ embedded in an elastic solid domain $\Omgs$ driven by boundary conditions applied on the outer part of the solid boundary $\pp\Omgs$.

In COMSOL, we specify user-defined equations and boundary conditions in weak form using the PDE mathematics module, and we express all vector fields in Cartesian coordinates $(x,y,z)$. At the boundary $\pp\Omgf$, the local right-handed  orthonormal basis $\big\{\exs, \eys,\ezs\big\}$ is implemented using the built-in COMSOL tangent vectors $\mathtt{t1}$ and $\mathtt{t2}$ as well as the normal vector $\mathtt{n}$, all given in Cartesian coordinates. Boundary-layer fields (suberscript "0"), such as $\vvvwall_1$, $\vvvls_1$, and $\vvvds_1$, are defined on the boundary $\pp\Omgf$ only, and their spatial derivatives are computed using the built-in tangent-plane derivative operator $\mathtt{dtang}$. For example, in COMSOL we call the Cartesian components of $\vvvds_1$ for $\mathtt{vdX}$, $\mathtt{vdY}$, and $\mathtt{vdZ}$ and compute $\div \vvvds_1$ as $\mathtt{dtang(vdX,x)} + \mathtt{dtang(vdY,y)+dtang(vdZ,z)}$. The models are implemented in COMSOL using the following two-step procedure:\cite{Muller2015}

Step (1), first-order fields\cite{Muller2014, Ley2016}: For a given frequency $\omega$,  the driving first-order boundary conditions for the system are specified; the wall velocity $\vvvwall_1$ on $\pp\Omgf$ for the fluid-only model, and the outer wall displacement $\uuu_1$ on $\pp\Omgs$ for the solid-fluid model. Then, the first-order fields are solved; the pressure $p_1$ in $\Omgf$ using \eqsref{Helmholtz_p}{p1_bc}, and, if included in the model, the solid displacement $\uuu_1$ in the solid domain $\Omgs$. In particular, in COMSOL we implement $\pp^2_\perp \pI = (\een_\zeta\cdot\nablabf)^2\pI$ in \eqref{p1_bc} as $\mathtt{nx*nx*p1xx + 2*nx*ny*p1xy + \ldots}$.

Step (2), second-order fields\cite{Muller2014, Muller2015}:  Time averages $\frac12 \Re\big\{f^* g\big\}$ are implemented using the built-in COMSOL operator $\mathtt{realdot}$ as $\mathtt{0.5*realdot(f,g)}$. Moreover, in the boundary condition~\eqnoref{vl2_bc_final}, the normal derivative of $\vl_{1\perp}$ in $\AAA$ is rewritten as $\pp_\perp \vl_{1\perp}=\div\vvvl_1-\div \vvvls_1=\ii\kappa_0\omega p_1^0-\div \vvvls_1$ for computational ease, and the advective derivatives in $\AAA$ and $\BBB$, such as the term $\Re \big\{ \vvvdsC_1\!\scap\grad\vvvds_1\big\}\cdot\een_x$ in $\AAA\cdot\ex$, are computed as  $\mathtt{realdot(vdX,}$ $\mathtt{dtang(vdX,x))}$ + $\mathtt{realdot(vdY,}$ $\mathtt{dtang(vdX,y))}$ + $\mathtt{realdot(vdZ,}$  $\mathtt{dtang(vdX,z))}$.

All numerics were carried out on a workstation, Dell Inc Precision T3610 Intel
Xeon CPU E5-1650 v2 at 3.50 GHz with 128 GB RAM and 6 CPU cores.




\begin{table}[b]
\centering
\caption{\tablab{material_param} Material parameters at 25 C$^\circ$ used in the numerical modeling presented in \secsref{ExI}{ExII}.}
\begin{ruledtabular}
\begin{tabular}{llrl}
\textit{Water} \citep{Muller2014}: &    &    & \\
Mass density     & $\rho_0$  & 997.05  & kg~m$^{-3}$ \\
Compressibility     & $\kappa_0$ & 452  & TPa$^{-1}$  \\
Speed of sound  & $c_0$ & 1496.7  & m~s$^{-1}$ \\
Dynamic viscosity & $\etaO$ & 0.890 & mPa\,s \\
Bulk viscosity & $\etaBO$ & 2.485 & mPa\,s \\
\textit{Pyrex glass} \citep{Corning_Pyrex}:& \rule{0mm}{1.1em}   &    &  \\
Mass density     & $\rhoS$  & 2230  & kg~m$^{-3}$ \\
Speed of sound, longitudinal  & $\cL$ & 5592  & m~s$^{-1}$ \\
Speed of sound, transverse  & $\cT$ & 3424  & m~s$^{-1}$ \\
Solid damping coefficient     & $\GammaS$ & 0.001 &   \\
\end{tabular}
\end{ruledtabular}
\end{table}

\begin{figure*}
\centering
\includegraphics[width=\textwidth]{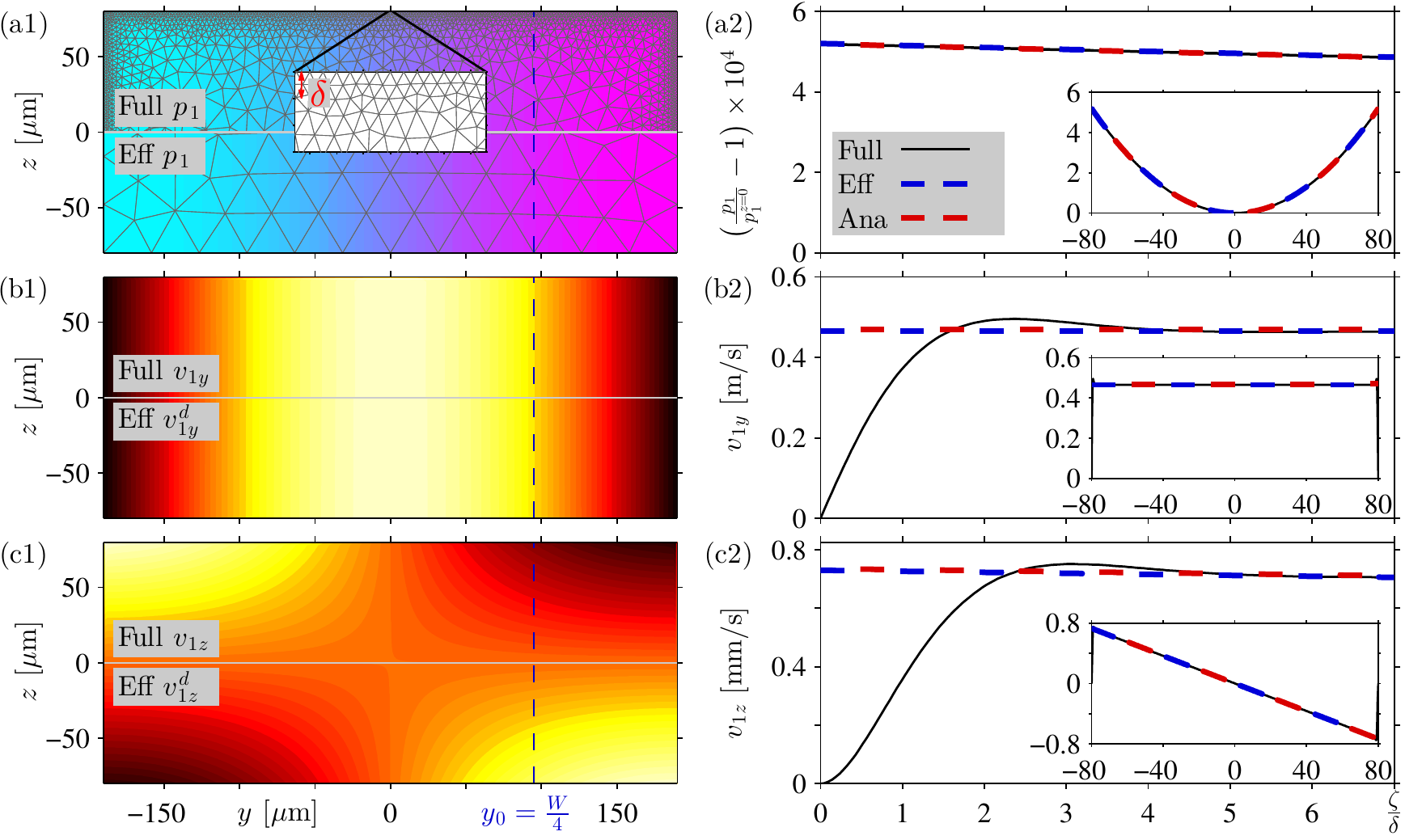}
\caption[]{\figlab{1st-order_rectangle}
First-order pressure and velocity fields in the vertical rectangular cross section of a long, straight channel of width $W = 380~\SImum$ and height $H = 160~\SImum$ at resonance $\fres = 1.967$~MHz. Color plots show the fields of the full (upper half) and effective (lower half) model for (a1) the pressure $p_1$ from $-1$ MPa (cyan) to 1 MPa (purple) and the finite element mesh (gray), (b1) the horizontal velocity $v_{1y}$ from 0~m/s (black) to 0.7~m/s (white), and (c1) the vertical velocity $v_{1z}$  from $-1$~mm/s (black) to 1~mm/s (white). Line plots at $y_0=\frac14 W$ for $-\frac12 H < z < -\frac12 H+ 7\delta$ (blue dashed line in the color plots) show (a2) the relative pressure deviation $p_1(y_0,z)/p_1(y_0,0)-1$, (b2) the horizontal velocity $v_{1y}$, and (c2) the vertical velocity $v_{1z}$. The insets show the respective line plots along the entire line $-\frac12 H < z < \frac12 H$. "Ana" refers to the analytical results from \eqref{first_fields_sol}.}
\end{figure*}

\section{Example I: A rectangular surface}
\seclab{ExI}
We apply our theory to a long, straight channel along the $x$ axis with a rectangular cross section in the vertical $y$-$z$ plane, a system intensively studied in the literature both theoretically \cite{Muller2012, Muller2014, Muller2015} and experimentally\cite{Barnkob2010, Augustsson2011, Barnkob2012, Muller2013}. We consider the 2D rectangular fluid domain $\Omgf$ with $-\frac12 W < y < \frac12 W$ and $-\frac12 H < z < \frac12 H$, where the top and bottom walls at $z = \pm\frac12 H$ are stationary and the vertical side walls at $y=\pm \frac12 W$ oscillate with a given velocity $\vwall_{1y}w(z)\eiot\eee_y$ and frequency $f=\frac{\omega}{2\pi}$ close to  $\frac{\cO}{2 W}$, thus exciting a half-wave resonance in the $y$-direction. In the simulations we choose the wall velocity to be $\vwall_{1y}=d_0\omega$ with a displacement amplitude $d_0=0.1~\SInm$. The material parameters used in the model are shown in \tabref{material_param}.

We compare the results from the effective theory with the full boundary-layer-resolved simulation developed  by Muller \etal\ \citep{Muller2012} Moreover, we derive analytical expressions for the acoustic fields, using pressure acoustics and our effective boundary condition \eqref{p1_bc}, and for the streaming boundary condition using \eqref{vl2_bc_final}.

\subsection{Pressure acoustics: First-order pressure}
To leading order~in~$\eps$ and assuming small variations in $z$, \eqsref{Helmholtz_p}{p1_bc} in the fluid domain $\Omgf$ becomes,
 \bsubal{rect_prob_ana}
 \lap p_1+k_0^2\: p_1 &=0,  &&\rrr \in \Omgf,
 \\
 \pp_y p_1 &=\ii \omega \rho_0 \vwall_{1y} w(z),
 && y=\pm \frac12 W, \eqlab{p_bc_y_rect}
 \\
 \mp \pp_z p_1 &= -\dfrac{\ii}{\ks}k_0^2\:p_1
 && z=\pm \frac12 H. \eqlab{p_bc_z_rect}
 \esubal
This problem is solved analytically by separation of variables, introducing $k_y$ and $k_z$ with $k_y^2+k_z^2=k_0^2$ and choosing a symmetric velocity envelope function $w(z)=\cos(k_z z)$. This leads to the pressure $p_1=A \sin(k_y y)\cos(k_z z)$, where $A$ is found from \eqref{p_bc_y_rect},
 \bal\eqlab{p1_sol_rect}
 p_1(y,z)=\dfrac{\ii\omega\rho_0 \vwall_{1y}}{k_y\cos(k_y\frac{W}{2})}\sin(k_y y)\cos(k_z z).
 \eal
According to \eqref{p_bc_z_rect}, $k_z$ must satisfy
 \bal
 k_0^2=\ii \ks k_z\tan\Big(k_z \dfrac{H}{2}\Big),
 \eal
and using $\tan(k_z \frac{H}{2})\approx \frac{1}{2} k_z H$ for $k_z H \ll 1$, we obtain
 \bal
 k_z^2= -(1+\ii)\frac{\delta}{H}\: k_0^2, \quad
 k_y^2= \Big[1+(1+\ii)\frac{\delta}{H}\Big]k_0^2.
 \eal
Note that $k_y$ becomes slightly larger than $k_0$ since the presence of the boundary layers introduces a small variation in the $z$ direction. The half-wave resonance that maximizes the amplitude of $p_1$ in \eqref{p1_sol_rect} is therefore found at a frequency $\fres$ slightly lower than $\fres^0 = \frac{c_0}{2W}$,
 \beq{fres_rect}
 \fres = \Big(1-\frac12\Gambl\Big)\:\fres^0 , \quad
 \text{ with }\;\Gambl = \frac{\delta}{H}.
 \eeq
Here, we introduced the boundary-layer damping coefficient $\Gambl$ that shifts $\fres$ away from $\fres^0$. This resonance shift is a result of the extended boundary condition~\eqnoref{p1_bc}, and it cannot be calculated using classical pressure acoustics.

Using $f=f_\mathrm{res}$ in \eqref{p1_sol_rect} and expanding to leading order in $\Gambl$, gives the resonance pressure and velocity,
 \bsubal{first_fields_sol}
 \frac{p_1^\mathrm{res}}{\rho_0 c_0} &=
 \frac{-4\vwall_{1y}}{\pi\Gambl}
 \bigg\{\!\sin(\yti)+\frac{\Gambl}{2}[\ii\yti\cos(\yti)-\sin(\yti)]\bigg\} Z_\mathrm{res}(\zti),
 \\
 \eqlab{first_fields_sol_b}
 v_{1y}^{d,\mathrm{res}}&=
 \frac{4\ii \vwall_{1y}}{\pi\Gambl}
 \bigg\{\!\cos(\yti)+\ii\frac{\Gambl}{2}\big[\cos(\yti)-
 \yti\sin(\yti)\big]\bigg\}Z_\mathrm{res}(\zti),
 \\
 v_{1z}^{d,\mathrm{res}}&=\frac{4\ii\vwall_{1y}}{\pi}(1+\ii)\sin(\yti)\zti,
 \esubal
where $\yti=\pi \frac{y}{W}$, $\zti= \pi \frac{z}{W}$, and $Z_\mathrm{res} = \big[1+\frac12\Gambl(1+\ii)\zti^2\big]$. Note that at resonance, the horizontal velocity component is amplified by a factor $\Gambl^{-1}$ relative to the wall velocity, $v_{1y}^{d,\mathrm{res}} \sim \Gambl^{-1} v_{1z}^{d,\mathrm{res}} \sim \Gambl^{-1} \vwall_{1y}$, while the horizontal component is not.

In \figref{1st-order_rectangle}, we compare an effective ("Eff") pressure-acoustics simulation of $\pI$ solving \eqsref{Helmholtz_p}{p1_bc}, with a full pressure-velocity simulation of $\pI$ and $\vvvI$ from \eqref{cont_navier_first} as in Muller and Bruus\citep{Muller2012}. The analytical  results ("Ana") for $p_1^\mathrm{res}$, $v_{1y}^{d,\mathrm{res}}$, and $v_{1z}^{d,\mathrm{res}}$ in \eqref{first_fields_sol} are also plotted along the line $y=\frac14 W$ in \figref{1st-order_rectangle}(a2), (b2), and (c2), respectively. The relative deviation between the full and effective fields outside the boundary are less than 0.1\% even though the latter was obtained using only $5\times 10^3$ degrees of freedom (DoF) on the coarse mesh compared to the $6\times 10^5$ DoF necessary in the former on the fine mesh. Note that from the effective model, the boundary-layer velocity field $\vvvd_1$ can be computed using \eqsref{vd_sol}{vvvds_sol}.

 \begin{figure}[!t]
 \centering
 \includegraphics[width=0.95\columnwidth]{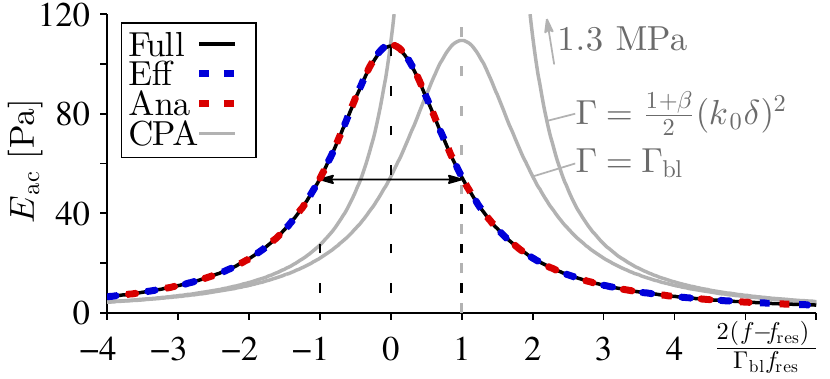}
 \caption[]{\figlab{EacResonance}
 Resonance curves for the rectangular channel. "Ana" refers to the analytical result from \eqref{Eac_b} and "CPA" refers to simulations using classical pressure acoustics with the boundary condition $\pp_\perp p_1=\ii\omega\vwallperp$ at $\rrr\in \partial \Omega$ with different choices of bulk damping coefficient $\Gamma$.}
 \end{figure}

To study the resonance behaviour of the acoustic resonator further, we compute the space- and time-averaged energy density $\avr{\Eacbar}$ stored in the acoustic field for frequencies $f$ close to the resonance frequency  $\fres$. Inserting $k_y=\frac{\pi}{W}(1+\frac{\ii}{2}\Gambl)+\frac{2\pi}{c_0}(f-\fres)$ into \eqref{p1_sol_rect}, results in the Lorentzian line-shape for $\avr{\Eacbar}$,
 \bsub
 \bal
 \eqlab{Eac}
 \nn
 \avr{\Eacbar}&= \avr{\bar{E}_\mr{ac}^{d,\mr{kin}}} + \avr{\bar{E}_\mr{ac}^{d,\mr{pot}}}
 = 2 \avr{\bar{E}_\mr{ac}^{d,\mr{pot}}}= 2 \avr{\bar{E}_\mr{ac}^{d,\mr{kin}}}
 \\
 &= \dfrac{2}{HW}\iint_\Omgf \dfrac{1}{2} \kapO \avr{p_1p_1}\ \dm y \dm z
 \\
 &\approx \dfrac{1}{\pi^2}\rho_0 \big(\vwall_{1y}\big)^2  \eqlab{Eac_b}
 \frac{1}{\big(\frac{f}{\fres}-1\big)^2+\big(\frac12\Gambl\big)^2},
 \text{ for } f\approx \fres.
 \eal
 \esub
From this follows the maximum energy density at resonance, $\avr{\Eacbarres} = \avr{\Eacbar(\fres)}$,  and the quality factor $Q$,
 \bal
 \avr{\Eacbarres}=\frac{1}{4} \rho_0 \left(\frac{4\vwall_{1y}}{\pi\Gambl}\right)^2, \qquad Q=\frac{1}{\Gambl}=\frac{H}{\delta}.
 \eal
As shown in \figref{EacResonance}, there is full agreement between the effective pressure-acoustics model, the full pressure-velocity model, and the analytical model. This is in agreement with the Q-factor obtained from \eqref{Q-factor},
 \beq{Q-factor-rect}
 Q=\frac{\displaystyle 2\iint_{\Omgf} \frac{1}{4}\rho_0 |v_{1y}^\mr{res}|^2\ \dm y\dm z}{
 \displaystyle 2\int_{-\frac{W}{2}}^{+\frac{W}{2}} \frac{1}{4}\delta\rho_0 |v_{1y}^\mr{res}|^2\ \dm y}
 =\frac{H}{\delta},
 \eeq
and also in agreement with the results obtained by Muller and Bruus\cite{Muller2015} and by Hahn \etal\cite{Hahn2015} using the approximation $\Ploss \approx \Pdiss$ in \eqref{Q-factor}.

\begin{figure}[t!]
\centering
\includegraphics[width=\columnwidth]{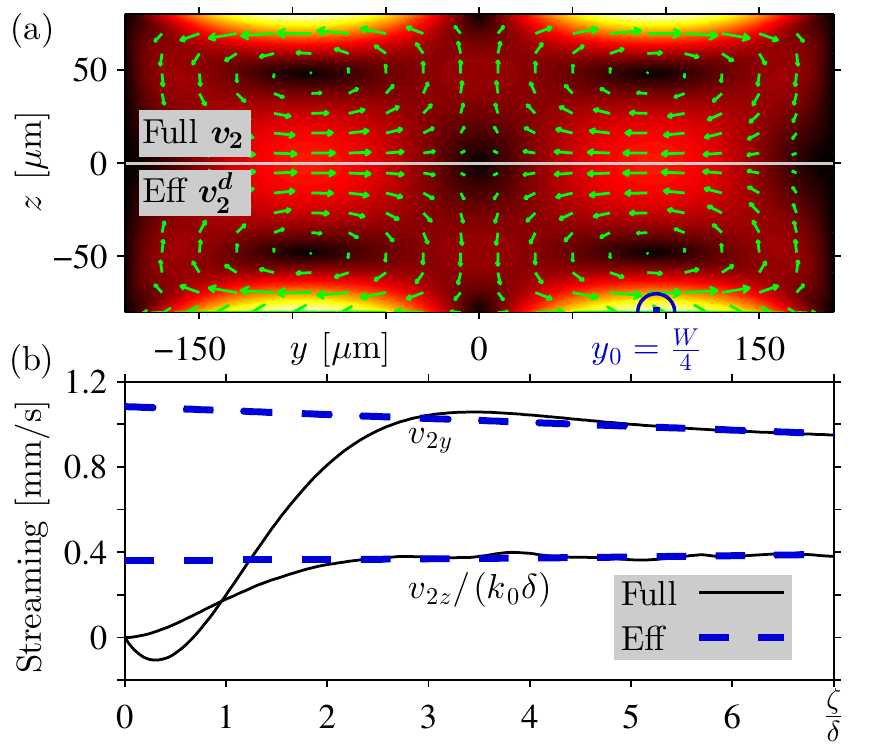}
\caption[]{\figlab{2nd-order_rectangle}
Second-order velocity for the rectangular channel. (a) Comparison of full $\vvv_2$ (above) and effictive (below) streaming $\vvv_2^d$. (b) Line plots at $y_0=\frac14 W$ for $-\frac12 H < z < -\frac12 H+ 7\delta$ near the center of the blue half circle in (a).}
\end{figure}

\subsection{Second-order streaming solution}
For the full model at resonance $\fres$, we solve \eqref{cont_navier_second}, while for the effective model we solve \eqref{cont_navier_second_long_simple} with the boundary condition on $\vvvl_2$ obtained by inserting the velocity fields from \eqref{first_fields_sol} into \eqref{vl2_bc}. At the surfaces  $z=\pm \frac12 H$, we find to lowest order in $\eps$,
 \bsubal{vl2_bcy_rect}
 \eqlab{vl2_bcy_rect_a}
 \vls_{2y} &= \frac{3}{8c_0}\bigg(\frac{4\vwall_{1y}}{\pi\Gambl}\bigg)^2\sin(2 \yti),
 \\
 \eqlab{vl2_bcy_rect_b}
 \vls_{2z} &= \mp(k_0\delta)\frac{1}{8c_0}\bigg(\frac{4\vwall_{1y}}{\pi\Gambl}\bigg)^2\big[1+10\cos(2\yti)\big].
 \esubal
The resulting fields of the two models are shown in \figref{2nd-order_rectangle}. Again, we have good quantitative agreement between the two numerical models, now better than 1\% or $3k_0\delta$, for $9\times 10^3$ DoF and $6\times 10^5$ DoF, respectively.

Analytically, \eqref{vl2_bcy_rect_a} is the usual parallel-direction boundary condition for the classical Rayleigh streaming \citep{LordRayleigh1884}, while \eqref{vl2_bcy_rect_b} is beyond that, being the perpendicular-direction boundary condition on the streaming, which is a factor $k_0\delta \approx 3\times 10^{-3}$ smaller than the parallel one. This is confirmed in \figref{2nd-order_rectangle}(b) showing the streaming velocity close to $z=-\frac12 H$ at $y=\frac14 W$.

\begin{figure}[b!]
\centering
\includegraphics[width=\columnwidth]{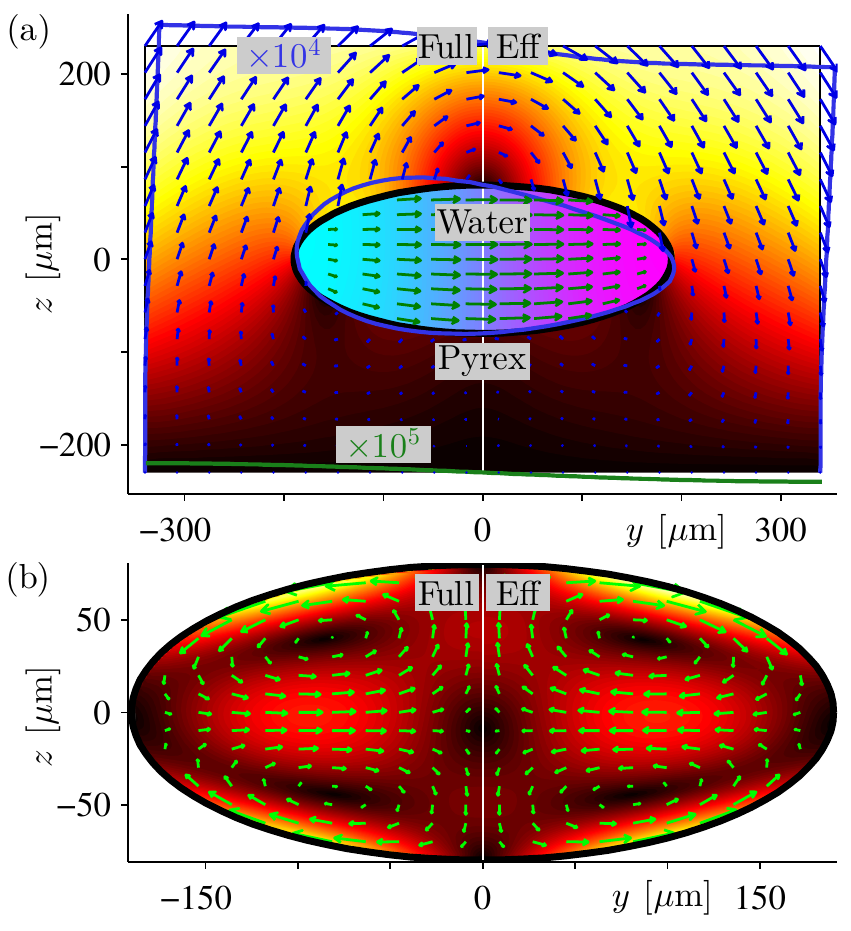}
\caption[]{\figlab{1st-2nd-order_ellipse}
Full (left) and effective (right) solutions for a curved channel with fluid-solid coupling. (a) Elliptic fluid domain with the acoustic pressure $p_1$ from $-0.3$ MPa (cyan) to $+0.3$ MPa (purple) and fluid velocity (green arrows, max $0.2$ m/s) surrounded by solid pyrex with displacement field $\uuu_1$ (blue arrows) and displacement magnitude $\abs{\uuu_1}$ from $0$ nm (black) to $2.7$ nm (yellow). To be visible, the displacement (blue line and blue arrows, max 3~nm) is enhanced $10^4$ times, except at the bottom (green line, max 0.1~nm) where it is enhanced $10^5$ times. (b)  Streaming velocity $\vvv_2$ (green arrows) and magnitude from $0$ $\SImum$/s (black) to $7.8~\SImum$/s (yellow).}
\end{figure}

\section{Example II: a curved oscillating surface}
\seclab{ExII}
Next, we implement in COMSOL our the boundary conditions \eqsref{p1_bc}{vl2_bc_final} in a system with a curved solid-fluid interface that oscillates in any direction, as described in \secref{Comsol_model}. We consider an ellipsoidal fluid domain (water) of horizontal major axis $W=380~\SImum$ and vertical minor axis $H=160~\SImum$ surrounded by a rectangular solid domain (Pyrex) of width $W_\mathrm{s}=680~\SImum$ and height $H_\mathrm{s}=460~\SImum$. We actuate the solid at its bottom surface using a velocity amplitude $\vwallperp=d_0\omega \sin(\frac{\pi y}{W_\mathrm{s}})$ with $d_0=0.1~\SInm$ and at the resonance frequency $\fres = 2.222$~MHz, which has been determined numerically as in \figref{EacResonance}. The governing equations for the displacement field $\uuu_1$ of the solid are those used by Ley and Bruus\citep{Ley2017},
 \bsubal{solid}
 \div\sigmat_\mathrm{s}&=-\rho_\mathrm{s}\omega^2 (1+\ii\Gamma_\mathrm{s})\uuu_1,  \; \text{ in the solid domain}\\
 -\ii\omega \uuu &= \vvvwall_1(y), \; \text{ actuation at } z=-\frac{1}{2}H_\mathrm{s},\\
 \nnn_\mathrm{s}\scap\sigmat_\mathrm{s}&=\zerovec, \; \text{ at solid-air interfaces},\\
 \nnn_\mathrm{s}\scap\sigmat_\mathrm{s}&=\nnn_{s}\scap\sigmat_1, \; \text{ at solid-fluid interfaces},
 \esubal
where $\sigmatS=\rhoS \cT^2[\grad\uuu+(\grad\uuu)^T]+ \rhoS(\cL^2-2\cT^2)(\grad\cdot\uuu)\Imat$ is stress tensor of the solid with mass density $\rhoS$, transverse velocity $\cT$, longitudinal velocity $\cL$, and damping coefficient $\Gamma_\mathrm{s}$, while $\nnn_\mr{s}$ is the solid surface normal, and $\nnn_\mr{s}\cdot\sigmat_1 = \ezs\cdot\sigmat_1$ is the fluid stress on the solid, \eqref{stress1_bc}. The material parameter values are listed in \tabref{material_param}.

We solve numerically \eqsref{Helmholtz_p}{p1_bc} in first order and \eqsref{cont_navier_second_long_simple}{vl2_bc_final} in second order. The results are shown in \figref{1st-2nd-order_ellipse}, where we compare the simulation results from the full boundary-layer resolved simulation of \eqref{cont_navier_second} with the effective model. Even for this more complex and realistic system consisting of an elastic solid with a curved oscillating interface coupled to a viscous fluid, we obtain good quantitative agreement between the two numerical models, better than $6\times 10^5$ DoF and 1\% for $9\times 10^3$ DoF, respectively.

\section{Conclusion}
We have studied acoustic pressure and streaming in curved elastic cavities having time-harmonic wall oscillations in any direction. Our analysis relies on the condition that both the surface curvature and wall displacement are sufficiently small as quantified in \eqref{lengthscale_assumptions}.

We have developed an extension of the conventional theory of first-order pressure acoustics that includes the viscous effects of the thin acoustic boundary layer. Based on this theory, we have also derived a slip-velocity boundary condition for the steady second-order acoustic streaming, which allows for efficient computations of the resulting incompressible Stokes flow.

The core of our theory is the decomposition of the first- and second-order fields into long- and short-range fields varying on the large bulk length scale $d$ and the small boundary-layer length scale $\delta$, respectively, see \eqsref{Navier_decomp}{decomp_second}. In the physically relevant limits, this velocity decomposition allows for analytical solutions of the boundary-layer fields. We emphasize that in contrast to the conventional second-order matching theory of inner solutions in the boundary layer and outer solutions in the bulk, our long- and short-range, second-order, time-averaged fields co-exist in the boundary layer; the latter die out exponentially beyond the boundary layer leaving only the former in the bulk.

The main theoretical results of the extended pressure acoustics in \secref{first_theory} are the
boundary conditions~\eqnoref{p1_bc} and~\eqnoref{stress1_bc} for the pressure $\pI$ and the stress $\sigmat_1\cdot\ezs$ expressed in terms of the pressure $\pI$ and the velocity $\vvvwall_1$ of the wall. These boundary conditions are to be applied to the governing Helmholtz equation~\eqnoref{Helmholtz_p} for $\pI$,  and the gradient form~\eqnoref{p_to_vl} of the compressional acoustic velocity field $\vvvl_1$. Furthermore, in \secref{Acoustic_power_loss}, we have used the extended pressure boundary condition to derive an expression for the acoustic power loss $\Ploss$, \eqref{PlossSuper}, and the quality factor $Q$, \eqref{Q-factor}, for acoustic resonances in terms of boundary-layer and bulk loss mechanisms. The main result of the streaming theory in \secref{second_theory} is the governing incompressible Stokes equation~\eqnoref{cont_navier_second_long_simple} for the streaming velocity $\vvvl_2$ and the corresponding extended boundary condition~\eqnoref{vl2_bc_final} for the streaming slip velocity $\vvvls_2$. In this context, we have developed a compact formalism based on the $I^{(n)}_{ab}$-integrals of \eqref{IabnEqs} to carry out with relative ease the integrations that lead to the analytical expression for $\vvvls_2$. Lastly, in \secref{Special_cases}, we have applied our extended pressure-acoustics theory to several special cases. We have shown, how it leads to predictions that goes beyond previous theoretical results in the literature by Lord Rayleigh~\cite{LordRayleigh1884}, Nyborg~\citep{Nyborg1958}, Lee and Wang~\citep{Lee1989}, and Vanneste and B\"{u}hler~\citep{Vanneste2011}, while it does agree in the appropriate limits with these results.

The physical interpretation of our extended pressure acoustics theory may be summarized as follows: The fluid velocity $\vvvI$ is the sum of a compressible velocity $\vvvl_1$ and an incompressible velocity $\vvvd_1$, where the latter dies out beyond the boundary layer. In general, the tangential component $\vvvwallpar = \vvvlspar + \vvvdspar$ of the no-slip condition  at the wall induces a tangential compression of $\vvvd_1$ due to the tangential compression of $\vvvl_1$ and $\vvvwall_1$. This in turn induces a perpendicular velocity component $\vdsperp$ due to the incompressibility of $\vvvd_1$. To fulfil the perpendicular no-slip condition $\vwallperp = \vlsperp + \vdsperp$, the perpendicular component $\vlsperp$ of the acoustic velocity must therefore match not only the wall motion $\vwallperp$, as in classical pressure acoustics, but the velocity difference $\vwallperp-\vdsperp$. Including $\vdsperp$ takes into account the power delivered to the acoustic fields due to tangential wall motion and the power lost from the acoustic fields due to tangential fluid motion. Consequently, by incorporating into the boundary condition an analytical solution of $\vvvd_1$, our theory subsequently leads to the correct acoustic fields, resonance frequencies, resonance Q-factors, and acoustic streaming.

In \secsrangeref{Comsol_model}{ExII} we have demonstrated the implementation of our extended acoustic pressure theory in numerical finite-element COMSOL models, and we have presented the results of two specific models in 2D: a water domain with a rectangular cross section and a given velocity actuation on the domain boundary, and a water domain with an elliptic cross section embedded in a rectangular glass domain that is actuated on the outer boundary. By restricting our examples to 2D, we have been able to perform the direct numerical simulations of the full boundary-layer-resolved model, and to use these results for validation of our extended acoustic pressure and streaming theory. Remarkably, we have found that our approach makes it possible to simulate acoustofluidic systems with a drastic nearly 100-fold reduction in the necessary degrees of freedom, while achieving the same quantitative accuracy, typically of order $k\delta$, compared to direct numerical simulations of the full boundary-layer resolved model. We have identified three reasons for this reduction: (1)~Neither our first-order nor our second-order method involve the fine-mesh resolution of the boundary layer. (2) Our first-order equations~\eqsnoref{Helmholtz_p}{p1_bc} requires only the scalar pressure $\pI$ as an independent variable, while the vector velocity $\vvvI$ is subsequently computed from $\pI$, \eqref{p_to_vl}. (3) Our second-order equations~\eqsnoref{cont_navier_second_long_simple}{vl2_bc_final} avoid the numerically demanding evaluation in the entire fluid domain of large terms that nearly cancel, and therefore our method requires a coarser mesh compared to the full model, also in the bulk.

The results from the numerical examples in \secsref{ExI}{ExII} show that the extended pressure acoustics theory has the potential of becoming a versatile and very useful tool in the field of acoustofluidics. For the fluid-only rectangular domain in \secref{ExI}, we showed how the theory not only leads to accurate numerical results for the acoustic fields and streaming, but also allows for analytical solutions, which correctly predict crucial details related to viscosity of the first-order acoustic resonance, and which open up for a deeper analysis of the physical mechanisms that lead to acoustic streaming. For the coupled fluid-solid system in 2D of an elliptical fluid domain embedded in a rectangular glass block, we showed in \secref{ExII} an important example of a more complete and realistic model of an actuated acoustofluidic system. The extended pressure acoustics theory allowed for calculations of acoustic fields and streaming with a relative accuracy lower than 1\%. Based on preliminary work in progress in our group, it appears that the extended pressure acoustic theory makes 3D simulations feasible within reasonable memory consumptions for a wide range of microscale acoustofluidic systems such as fluid-filled cavities and channels driven by attached piezoelectric crystals as well as droplets in two-phase systems and on vibrating substrates.

Although we have developed the extended pressure-acoustics theory and corresponding streaming theory within the narrow scope of microscale acoustofluidics, our theories are of general nature and may likely find a much wider use in other branches of acoustics.

\appendix
\section{Acoustic power balance}\seclab{Full_E_balance}
The time averages  $\avr{\Eac^\mr{kin}}$, $\avr{\Eac^{\mr{pot}}}$, and $\avr{\Eac}$ of the kinetic, the potential, and the total acoustic energy density, respectively, are given by
 \bsubal{EkinAVR}
 \avr{\Eackin}	&=\frac{1}{2}\rho_0\avr{\vvv_1\scap\vvv_1},
 \\
 \avr{\Eacpot} &= \frac{1}{2}\kappa_0\avr{p_1p_1},
 \\
 \avr{\Eac}&=\avr{\Eackin}+\avr{\Eacpot}.
 \esubal
Using Gauss's theorem and $\rhoO\pp_t\vvvI = \div\sigmat_1$, the time-averaged total power delivered by the surrounding wall is written as the sum of the time-averaged rate of change of the acoustic energy and total power dissipated into heat,
 \bsubal{E_balance}
 \oint_{\pp \Omega} &\avr{\vvvwall_1\scap\sigmat_1}\scap\nnn \, \dm A
 =\int_\Omega \div \avr{\vvv_1\scap \sigmat_1} \, \dm V
 \\
 &=\int_\Omega \Big[\avr{\vvv_1\scap (\div \sigmat_1)}+\avr{(\grad \vvv_1)\scapmat\sigmat_1}\Big] \dm V,
 \\
 \eqlab{E_balance_final}
 &=\int_\Omega \Big[\avr{\ppt \Eac}+\avr{(\grad \vvv_1)\scapmat\taut_1}\Big] \dm V.
 \esubal
Solving for the time-averaged change in acoustic energy $\int_\Omega \avr{\ppt \Eac} \, \dm V $ in \eqref{E_balance_final} gives
 \bsubal{Eac_change}
 &\int_\Omega \avr{\ppt \Eac} \, \dm V
 =\oint_{\pp \Omega} \avr{\vvvwall_1\scap\sigmat_1}\scap\nnn \, \dm A
 -\int_\Omega \avr{(\grad \vvv_1)\scapmat\taut_1}\, \dm V
 \\
 \eqlab{Eac_changeb}
 &\qquad =\oint_{\pp \Omega} \avr{\vvvwall_1(-\pI)}\scap\nnn \, \dm A
 +\int_\Omega \avr{\vvv_1\scap(\grad\scap\taut_1)}\, \dm V,
 \esubal
where Gauss's theorem transforms $\int_{\pp\Omega} \avr{\vvvwall_1\scap\taut_1}\scap\nnn\, \dm A$ into a volume integral, and $\nnn=-\ezs$ is the normal vector of the fluid domain $\Omega$. We may interpret \eqref{Eac_changeb} as the rate of change of stored energy in terms of a power $\avr{\Pvisc}$ due to viscous effects,
 \bal
  \avr{\ppt \Eac} = \avr{\Pvisc} = \avr{\Pdiss}+\avr{\Psurr},
 \eal
where $\avr{\Pdiss}$ is the viscous power dissipation into heat, and $\avr{\Psurr}$ is the power from the viscous part of the work performed by the wall on the fluid,
 \bsubal{Pdiss_Ploss}
  \avr{\Pvisc}&=\int_\Omega \avr{\vvv_1\scap(\grad\scap\taut_1)} \ \dm V,\\
 \avr{\Pdiss}&=-\int_\Omega \avr{(\grad \vvv_1)\scapmat\taut_1} \ \dm V,\\
 \avr{\Psurr}&=\oint_{\pp\Omega} \avr{\vvv_1\scap\taut_1}\scap \nnn \ \dm A.
 \esubal
Using \eqsref{Helmholtz_decomp}{Navier_decomp} we can evaluate $\avr{\Pvisc}$,
 \bsubal{Pvisc}
 \avr{\Pvisc}&=\int_\Omega \avr{\vvv_1\scap(\grad\scap\taut_1)}\, \dm V
 \\
 &=\int_\Omega \avr{\vvv_1\scap\big(\ii\Gamma\grad p_1-\ii\omega\rho_0\vvvd_1\big)}\: \dm V
 \\
 \nn
 &= \int_\Omega \bigg[-\dfrac{\Gamma\omega\rho_0}{2} |\vvvl_1|^2
 +\avr{\ppt \Eac^{\mr{kin},\delta}}\bigg] \dm V
 \\
 \eqlab{Pvisc_last}
 &\hspace{5mm}-\oint_{\pp\Omega} \avr{p_1\vvvds_1}\scap\nnn \, \dm A,
 \esubal
where we used \eqref{Navier_decomp} and Gauss' theorem. Inserting \eqref{Pvisc_last} into \eqref{Eac_changeb} leads to \eqref{global_energy_rate}. Comparing with \eqref{PlossSuper}, we can relate $\avr{\Ploss}=\avr{\Ploss^d}$ and $\avr{\Pvisc}$,
 \bsubal{Ploss_vs_Pvisc}
 &\avr{\Ploss}=\avr{\Pvisc}-\oint_{\pp\Omega} \avr{p_1\Big[\frac{\ii}{\ks}\pardiv\vvvwallpar\Big]}\scap\nnn \, \dm A\\
 &=\avr{\Pdiss}+\avr{\Psurr}-\oint_{\pp\Omega} \avr{p_1\bigg[\frac{\ii}{\ks}\pardiv\vvvwallpar\bigg]}\scap\nnn \, \dm A.
 \esubal
Note that $\avr{\Ploss}$ is not in general the same as the power $\avr{\Pdiss}$ dissipated into heat. These might however be approximately equal if the power $\oint_{\pp \Omega} -\avr{p_1\vvvwall_1}\scap\nnn \, \dm A$ delivered by the pressure is approximately balanced by dissipation $\avr{\Pdiss}$. This happens, if $\oint_{\pp \Omega} -\avr{p_1\vvvwall_1}\scap\nnn \, \dm A$ is much larger than $\avr{\Psurr}$ and $\oint \avr{p_1\Big[\frac{\ii}{\ks}\pardiv\vvvwallpar\Big]}\scap\nnn \, \dm A$, which is usually satisfied.

%
%


%

\end{document}